\documentclass[11pt,a4paper]{article}

\usepackage{jheppub}
\usepackage{amssymb,amsmath,amsthm,graphicx}
\usepackage{amssymb,comment,array,ulem}
\usepackage{caption}
\usepackage{subcaption}
\usepackage{bm}
\usepackage{orcidlink}

\newcolumntype{M}[1]{>{\hbox to #1\bgroup\hss$}l<{$\egroup}}
\makeatletter
\newcommand\@brcolwidth{3.67em}
\newenvironment{prmatrix}{%
    \left(%
    \hskip-\arraycolsep
    \new@ifnextchar[\@brarray{\@brarray[\@brcolwidth]}%
}{%
    \endarray
    \hskip -\arraycolsep
    \right)%
}
\def\@brarray[#1]{\array{r*\c@MaxMatrixCols {M{#1}}}}
\makeatother

\title{Hot entanglement? -- Parametrically coupled quantum oscillators in two heat baths: instability, squeezing and driving}
\author[a]{Onat Ar{\i}soy\orcidlink{0000-0002-3736-2369},}
\author[b]{Jen-Tsung Hsiang\orcidlink{0000-0002-9801-208X}}
\author[c]{and Bei-Lok Hu\orcidlink{0000-0003-2489-9914}}

\affiliation[a]{Chemical Physics Program, Institute for Physical Science and Technology,  University of Maryland, College Park, Maryland 20742, USA}
\affiliation[b]{Center for High Energy and High Field Physics, National Central University, Taoyuan 320317, Taiwan, ROC}
\affiliation[c]{Maryland Center for Fundamental Physics and Joint Quantum Institute,  University of Maryland, College Park, Maryland 20742, USA}

\emailAdd{}
\emailAdd{cosmology@gmail.com}
\emailAdd{blhu@umd.edu}
\abstract{Entanglement being a foundational cornerstone of quantum sciences and the primary resource in quantum information processing, understanding its dynamical evolution in realistic conditions is essential. Unfortunately, numerous model studies show that degradation of entanglement from a quantum system's environment, especially thermal noise, is almost unavoidable. Thus the appellation `hot entanglement' appears like a contradiction, until Galve et al [Phys. Rev. Lett. 105 180501 (2010)] announced that  entanglement can be kept at high temperatures if one considers a quantum system with time-dependent coupling between the two parties, each interacting with its individual bath. With the goal of understanding the sustenance of entanglement at high temperatures, working with the same model and set up as Galve et al, namely, parametrically-driven coupled harmonic oscillators interacting with their own Markovian baths,  this work probes into the feasibility of `hot entanglement' from three aspects listed in the subtitle. Our findings show that 1) hot entanglement functions only in the unstable regimes, 2) instability is a necessary but not sufficient condition, and 3) the power intake required by the drive operating in the unstable regime to sustain entanglement increases exponentially. The last factor indicates that hot entanglement under this modeling is theoretically untenable and its actual implementation likely unattainable.}  

\keywords{xxx, yyy}
\begin{document}

\maketitle


\allowdisplaybreaks
\baselineskip=18pt

\section{Introduction}

Quantum entanglement is an important theme of current research in theoretical and applied sciences because at the level of foundational theory it is the ``uniquely distinct feature of quantum"  \cite{Schrodinger,Schrodinger2} and, as the primary resource of quantum information processing, it acts as the fountain spring for quantum  computing, communication, control, cryptography and metrology  which usher in the so-called second revolution of quantum sciences and engineering we are witnessing now. Unfortunately, when considered under realistic physical conditions,  a system unavoidably interacts with its many environments, and model studies (see, e.g., \cite{YuEberly,AEPW02,FicTan06,ASH06,ASH09,Goan,AnZhang,Mani,Mani2,LCH08,LinHu09,entrev-horodecki,Ludwig,entg-book1,entg-book2,Kanu,Wilson,HH15AOP,HH15PRD,HH15PLB,HH15JHEP} and references therein) have found that entanglement in a realistic system can easily be degraded by various noises in the environments. It is well-known that thermal noise is the nemesis of many quantum features, quantum coherence and entanglement are no exception.  Thus it came somewhat as a surprise when Galve et al \cite{galve-prl}  announced their interesting finding that  entanglement can be kept at high temperature (`hot entanglement'~\cite{Vedral}) if one considers a parametrically driven quantum system with time-dependent coupling between the two parties, each interacting with its own bath. This is the theme we shall focus on here and probe deeper into.  A related noteworthy paper  by Estrada and Pach\'on \cite{EstPac} explores nonMarkovian effects on hot entanglement in the same setup as Galve et al by introducing a finite cutoff frequency in the baths. They show that non-Markovian dynamics, when compared to the Markovian case, allow for the survival of stationary entanglement at higher temperatures, with larger coupling strength to the baths and at smaller driving rates. However, neither of these two important works considered the work cost of the driving protocol with which the entanglement is sustained.  

Let us begin our story with some background on hot entanglement. 

\subsection{Background on `hot entanglement'}
    
The observation of quantum entanglement in hot and/or large scale systems has been of interest with different motivations. The relevance of entanglement in the context of quantum computation and information (see Refs.~\cite{densecoding,entg-dist1} as earlier examples of works relating entanglement distillation to quantum dense coding, communication, Ref.~\cite{entg-dist2} for the role of entanglement in quantum error correction, Ref.~\cite{qcomp-exp-spup1} necessity of entanglement for exponential speedup in quantum computation with pure states) and the challenges to sustain entanglement over times much larger than the decoherence timescale need to be mentioned as one of the major motivations behind many of the works on hot entanglement (see Ref.~\cite{qcomp-hotentg1} for hot environment mediated entanglement between solid state spin qubits, Ref.~\cite{qcomp-hotentg2} for entanglement between two-level atoms in an optical cavity and Ref.~\cite{qcomp-negtemp-entg} for entanglement with negative temperature Markovian baths). Other important directions in the research on hot entanglement include the relation between the heat current and entanglement between two qubits coupled to Markovian reservoirs at different temperatures~\cite{qcomp-entg-flow}, non-linear Hamiltonians, bath spectrum filtering and coupling via an auxiliary system~\cite{qcomp-entg-aux}. One other important field in which the entanglement between macroscopic objects at hot temperatures is relevant is the relatively new field of quantum biology where the quantum nature of molecules with hundreds of atoms is still under debate (see Ref.~\cite{qbio-entg-dyn} for an example of the interplay between the classical and quantum aspects of biomolecules in the context of driving induced entanglement in hot environments) and this ongoing debate is promising to open new horizons in molecular biology such as, but not limited to, enhanced biological measurements with quantum spectroscopy using entangled photons~\cite{qbio-entg-meas}, the effects of entanglement in photosynthetic light harvesting complexes on the efficiency of excitation transfer (see Ref.~\cite{qbio-entg-lhc} and references therein), chemical compass in birds~\cite{qbio-entg-comp1,qbio-entg-comp2}, electron tunneling assisted by phonon degrees of freedom of an odorant as an accurate model of olfaction~\cite{qbio-entg-olf1,qbio-entg-olf2} and possible effects on mutation rates~\cite{qbio-entg-mut}. Some other contexts in which entanglement is relevant are touched upon in Refs.~\cite{entrev-horodecki,entg-book1,entg-book2} and references therein.

\subsection{This work: Key issues}

This paper is a sequel of a 2015 paper by two of us \cite{HH15JHEP} on  the possibility of sustaining entanglement at  high temperatures between two coupled harmonic oscillators each interacting with its own bath at different temperatures, and related to a  recent paper \cite{HAH22} where we studied the effect of nonMarkovian baths on hot entanglement. Both studies assume constant inter-oscillator coupling. The conclusion drawn in \cite{HH15JHEP} is, in most realistic circumstances, for bosonic systems with \textit{constant} bilinear coupling interacting with  Ohmic (scalar field) baths, `hot entanglement' is unlikely. Similar qualitative conclusions were found in \cite{HAH22} for the same system with nonMarkovian baths. (In a later work \cite{AHH-nM} we shall address the nonMarkovian bath issues in comparison with Estrada and Pach\'on \cite{EstPac}). 

Here, we allow  the inter-oscillator coupling to change in time (driven parametric coupling) in the same set up as Galve et al, where the baths are Ohmic with infinite\footnote{{Here the Caldeira-Leggett approximation have been implemented because the high-temperature bath is considered.}} cutoff frequencies. At high temperature, the dynamics of the reduced system  is Markovian and the equations are a lot easier to solve than the nonMarkovian cases. 
Our analysis of the system's dynamics suggests that the parameters Galve et al used to report on hot entanglement fall in the dynamically unstable regimes. This is probably known to them, but they didn't probe into the special circumstances which allow for entanglement to be sustained and face the unwelcome consequences of dynamically unstable driven systems.

As we shall see there are many subtle issues, perhaps known, not widely,  but not addressed earlier (hot entanglement functions only in the unstable regimes) or hitherto unknown (instability is a necessary but not sufficient condition, {unreliability} of the Caldeira-Leggett approximation) and some concerning factors (costly power intake) which makes actual implementation of hot entanglement doubtful. 
	
\subsection{Major findings: Unstable dynamics, squeezed systems and energy budget}

Three key issues we wish to address here are: Q1) In What parameter regimes are hot entanglement functional? Does high temperature entanglement need be operated in a dynamically unstable regime?  Q2) Would squeezed systems without parametric drive allow for hot entanglement? Q3) If yes is the answer to Q1, would the energy budget needed for the parametric drive to sustain entanglement at high temperature be cost-efficient?  The answers to these questions are, put succinctly, 1) Yes, 2) No,  3) Not really.

1) \textit{Unstable dynamics}:  Our analysis suggests that the parameters  Galve et al use to report on hot entanglement fall in the dynamically unstable regimes. Whether the system can maintain a steady state needs be addressed.  Similar concerns as ours  have been expressed earlier, e.g., {Figueiredo Roque and Roversi} \cite{FigFov13} show that quantum correlations and entanglement in a system of two harmonic oscillators with time-dependent coupling and in contact with a {\it common} heat bath (note: this is different from our set-up) can survive  even at very high temperatures. They show that a`remarkable' relation exists between entanglement and the instability of the system, and that quantum correlations are much more sensitive to the parameters of the oscillators than the temperature of the bath.Chakraborty  and  Sarma \cite{ChaSar18} study the entanglement dynamics of two coupled mechanical oscillators within a modulated optomechanical system and identified  the critical mechanical coupling strength for transition from stationary to dynamical entanglement, which is extremely robust against the oscillator temperature. They show that this entanglement dynamics is strongly related to the stability of the normal modes. 

2) \textit{Squeezing}. {Unstable quantum (open) dynamics intrinsic in the system or due to external drive is expected to effectively induce large squeezing~\cite{GP85, GS86, HH22, GV20}. We are interested in whether the squeezing of a quantum system is sufficient to produce and sustain entanglement. A lesser known aspect in this context is that runaway dynamics also incites large thermal fluctuations. Survivability of quantum entanglement then can be understood as a tug of war between squeezing and thermal fluctuations. To show that squeezing is not sufficient condition we use an {amplifying harmonic (anti-damping)} oscillator interacting with a single bath to demonstrate that effective squeezing generated by dynamical instability is not strong enough to sustain entanglement.  Galve et al's result indicate that parametric drive is needed to boost squeezing and at the same time tame the stimulated thermal fluctuations. Thus instability does not guarantee entanglement: the unstable motion may induce a much larger effective thermal fluctuations than the effective squeezing can suppress. In contrast, for stable dynamics due to constant inter-oscillator coupling, as studied in \cite{HAH22}, squeezing in the initial state does not help late-time sustainability of entanglement because its benefit is severely weakened by the relaxation process.}

3) \textit{Power intake}.  
If it is confirmed that hot entanglement can exist only  in the unstable regime then the natural question to ask is, lacking a NESS, how long can one operate in such a regime and reap the benefits of sustained entanglement at high temperatures. For this inquiry we shall carry out an energy budget analysis to assess the power the external drive needs to deliver to the system to this end. It turns out to be unattainable, not entirely surprisingly for unstable systems.

This paper is organized as follows. In Sec.~\ref{S:ebtgdfg} we present the essential theoretical elements to treat quantum entanglement of the above-described parametrically-driven system. In Sec.~\ref{numer-ohmic} we compare the time evolution of hot entanglement when the system undergoes unstable and stable motion. In particular we focus on its sustainability at late times. In Sec.~\ref{S:eoujhbdf} we use the model of an amplifying harmonic oscillator as a counter example to illustrate the point that instability does not guarantee hot entanglement although it seems necessary. We introduce terms familiar in quantum optics -- effective squeezing and effective temperature -- to describe entanglement. Finally in Sec.~\ref{S:ebgdfhg}, we examine the energy flow between the external agent, the reduced system and the thermal bath. We arrive at the conclusion that hot entanglement under the present model and set-up \cite{galve-prl} does not seem tenable, by virtue of the fact that it needs to operate in the regime of unstable dynamics wherein a tremendous amount of energy consumption from the external agent is required.  Appendix~\ref{S:peotr} collects concise information about two-mode squeezing, needed for Sec.~\ref{S:eoujhbdf}. In Appendix~\ref{S:gbdghf}, we take a closer look at the Caldeira-Leggett approximation, applied to a high-temperature unstable system, and point out the ambiguities that are not present in stable dynamics.

\section{Parametrically coupled oscillators}\label{S:ebtgdfg}

\subsection{Theoretical framework}

It is physically transparent and mathematically simple to use the Langevin equations to describe the dynamics of the oscillator in this NESS setting~\cite{HH15AOP}. Since they are linear equations, we can readily write down its formal solutions from which we may construct the covariance matrix. This matrix is of central importance for the Gaussian system because it encompasses the essential features of the full dynamics~\cite{Peres, Horod, PPTSimon, Adesso, Adesso05, Illuminati04}. In particular, the quantum entanglement measure used here, logarithmic negativity, can be expressed in terms of the covariance matrix elements.

Suppose both oscillators have the same mass $m$, and physical frequency $\omega$ but we allow different oscillator-bath coupling strengths $e_{i}$ with $i=1$, 2. Let the displacements of the oscillators be denoted by $\chi_{i}$ with $i=1$, 2, and are grouped into a column vector $\bm{\chi}=(\chi_{1}\;\chi_{2})^{T}$. The Langevin equations for the current setup take on a compact form
\begin{align}
	\ddot{\bm{\chi}}(s) + 2\bm{\gamma} \cdot \dot{\bm{\chi}}(s) + \bm{\Omega}^2_{\textsc{r}}(s) \cdot \bm{\chi}(s) = \frac{1}{m}\, \bm{\xi}(s) \label{lange-sf}\,,
\end{align}
where $s$ is the time variable and the matrices $\bm{\gamma}$, $\bm{\Omega}^2_{\textsc{r}}(s)$ are given by
\begin{align}
	\bm{\gamma} &= \begin{pmatrix} \gamma_{1} &0  \\ 0 &\gamma_{2} \end{pmatrix}\,,&\bm{\Omega}^2_{\textsc{r}}(s) &=\begin{pmatrix} \omega^2 &\sigma(s) \\ \sigma(s) &\omega^2 \end{pmatrix}\, .
\end{align}
where $\gamma_{i}=e^2_i/(8\pi m)$ denotes the damping constant for oscillator $i$, and $\sigma(s)$ the time-dependent, inter-oscillator coupling strength.

On the righthand side \eqref{lange-sf}, $\bm{\xi}(s)$ is a Gaussian noise from the private baths, with the properties
\begin{align}
	\langle \bm{\xi}(s) \rangle &= 0, &\langle \bm{\xi}(s) \bm{\xi}^{T}(s') \rangle &= \begin{pmatrix} \nu^{(1)}(s,s') &0 \\  0 & \nu^{(2)}(s,s') \end{pmatrix} \label{xi-covar}\,.
\end{align}
It reflects the quantum fluctuations of the thermal baths the oscillators are separately attached to. In particular, the kernel function $\nu^{(i)}(s,s')=e_{i}^{2}G_{H,0}^{(i)}(s,s')$ contains the noise kernel $G_{H,0}^{(i)}(s,s')$ of the private bath $i$, and the coupling strength $e_{i}$. Thus the Langevin equations \eqref{lange-sf} describe the \textit{damped} dynamics of two parametrically coupled harmonic oscillators, driven by the quantum thermal noises from the their own baths. Here, although the equation contains a damping term, the evolution of such a system is not guaranteed to be stable. It could possess runaway solutions, depending on the choice of parametric driving. 

Following Ref.~\cite{HH15AOP} with generalization to the time-dependent frequency $\bm{\Omega}_{\textsc{r}}(s)$, the solution of the Langevin equation can be conveniently expressed in terms of the fundamental solution matrices $\bm{D}_1 (s,s')$ and $\bm{D}_2 (s,s')$ with the following conditions
\begin{align}
	\bm{D}_1 (s,s) &=1\,, &\dot{\bm{D}}_1 (s,s)&=0\,, &\bm{D}_2 (s,s) &= 0\,, &\dot{\bm{D}}_2 (s,s) &= 1\,,\label{d2-init}
\end{align}
and $\bm{D}_{i}(s,s')=0$ with $i=1$, 2 if $s<s'$. The matrices $\bm{D}_{i}$ have two time arguments due to the fact that the coefficients in the Langevin equation are time dependent, meaning that the response to the impulse depends on when the response is measured as well as when the impulse is applied. Thus it is not time translationally invariant, i.e., $\bm{D}_i(s,s')\neq\bm{D}_i(s-s')$. However, when the parametric driving is periodic, i.e., the matrix $\bm{\Omega}^2_{\textsc{r}}(s)$ being periodic, and the motion is stable, the fundamental solution matrices at late times have a nice property~\cite{paz-cool1}
\begin{align}\label{E:dkjgbfkse}
	\bm{D}_{i}(s,s') &= \bm{D}_{i} (s - n \tau_d, s' - n \tau_d)\,, &i&=1,2\,,
\end{align}
where $\tau_d$ is the driving period and $n$ is an integer.

Having defined the fundamental solution matrices, we find the displacement $\bm{\chi}$ and the corresponding canonical momentum $\bm{p}= m \dot{\bm{\chi}}(s) $ are given by
\begin{align}
	\bm{\chi}(s) &= \bm{D}_1 (s,0) \cdot \bm{\chi}(0) + \frac{1}{m}\,\bm{D}_2 (s,0) \cdot \bm{p}(0) + \frac{1}{m} \int_0^s ds'\; \bm{D}_2 (s,s') \cdot \bm{\xi}(s')\,, \label{chi} \\
	\bm{p}(s) &= m\dot{\bm{D}}_1 (s,0) \cdot \bm{\chi}(0) +\dot{\bm{D}}_2 (s,0) \cdot \bm{p}(0) +  \int_0^s ds' \;\dot{\bm{D}}_2 (s,s') \cdot \bm{\xi}(s')  \label{p}
\end{align}
where an overdot represents taking the derivative with respect to the first time arguments of the matrices $\bm{D}_i$.

The quantum Langevin equation (\ref{lange-sf}) of the coupled harmonic oscillators allows us to readily write down the evolution equations for the first moments of the canonical variables of the reduced system~\cite{qle-ford}
\begin{align}
	\frac{d \langle \chi_{i} \rangle}{dt} &= \frac{1}{m} \langle p_{i} \rangle \label{q-1st-mom-eqn} \\
	\frac{d \langle p_{i} \rangle}{dt} &= -2\gamma_{i} \langle p_{i}\rangle - m\omega^2 \langle \chi_{i} \rangle - m\sigma(t) \langle \chi_{j}  \rangle \label{p-1st-mom-eqn}
\end{align}
where the index $j=3-i$ with $i=1$, 2. This is a form of the Ehrenfest theorem. Notice that the noise term in the Langevin equation (\ref{lange-sf}) does not appear in Eqs.~\eqref{q-1st-mom-eqn} and \eqref{p-1st-mom-eqn} since the noise has zero mean, as assumed in \eqref{xi-covar}. For simplicity, we will assume that the means of the canonical variables are zero for the initial Gaussian state of the oscillators. Then, the equations of motion of the second moments take simpler forms~\cite{qle-ford}
\begin{align}
	\frac{d}{dt}\langle\chi_{i}^2\rangle&= \frac{1}{m} \langle \{  \chi_{i},\,p_{i}  \} \rangle\,, \\ 
	\frac{d}{dt}\langle \{\chi_{i},\, \chi_{j}\bigr\} \rangle&= \frac{1}{m} \langle\{ p_{i},\, \chi_{j}\}\rangle+ \frac{1}{m} \langle\{\chi_{i}, p_{j}\}\rangle\,, \\
	\frac{d}{dt}\langle\{ \chi_{i},\, p_{i}\}\rangle& =\frac{2}{m}\langle p_{i}^2 \rangle + \langle \{\xi_{i}-2\gamma_{i} p_{i} -m\omega^2 \chi_{i} -m\sigma(t)\chi_{j},\, \chi_{i}\rangle\,,    \\
	\frac{d }{dt}\langle\{ \chi_{i},\, p_{j}\}\rangle & = \frac{1}{m}\langle \{p_{i},\,p_{j}\}\rangle +\langle \{\xi_{j}-2\gamma_{j} p_{j} -m\omega^2 \chi_{j} -m\sigma(t)\chi_{i},\, \chi_{i}\} \rangle    \\
	\frac{d}{dt}\langle \{p_{i},\, p_{j}\}\rangle& =\langle \{\xi_{i}-2\gamma_{i} p_{i} -m\omega^2 \chi_{i} -m\sigma(t)\chi_{j},\, p_{j}\} \rangle\notag \\
		&\qquad\qquad\qquad\qquad\qquad+\langle \{\xi_{j}-2\gamma_{j} p_{j} -m\omega^2 \chi_{j} -m\sigma(t)\chi_{i},\, p_{i}\} \rangle\,,\\
	\frac{d}{dt}\langle p_{i}^2\rangle&= \langle \{\xi_{i}-2\gamma_{i} p_{i} -m\omega^2 \chi_{i} -m\sigma(t)\chi_{j},\, p_{i}\}\rangle\,, \label{E:ughdf}    
\end{align}
with $j=3-i$ and $i=1$, 2. These constitute a set of coupled differential equations for the second moments, i.e, covariance matrix elements, of the canonical variables for the coupled oscillator system, which are formally defined as~\cite{PPTSimon,Illuminati04,HH15JHEP}
\begin{equation}
	\bm{\sigma} =\frac{1}{2}\langle \{\bm{Z},\bm{Z}^{T} \} \rangle \,,
\end{equation}
by the phase space variale $\bm{Z}=(\chi_{1}\;\chi_{2}\;p_{1}\;p_{2})^{T}$. Finding the numerical solutions of the covariance matrix elements is both mathematically and computationally challenging in the parametrically driven open-system setting on account that both $\chi_{i}$ and $p_{i}$ are formally given by \eqref{chi} and \eqref{p}. However, at the high bath temperature regime, great simplification is possible by resorting to the Caldeira-Leggett limit~\cite{hm94} of the kernel function $\nu^{(i)}(s,s')$
\begin{equation}
	\nu^{(i)}(s,s') = \frac{4m\gamma_{i}}{\beta_{i}^{\textsc{b}}}\,\delta(s-s')\,,
\end{equation}
where $\beta_{i}^{\textsc{b}}$ is the initial inverse temperature of the $i$th private bath\footnote{There are a few subtleties when implementing the limit in the unstable system. {For example, other than the result given by the Caldeira-Leggett approximation, the $\langle\xi_ip_i\rangle$ in \eqref{E:ughdf} has a component which oscillates with exponentially increasing amplitude.}  We will elaborate them in  Appendix~\ref{S:gbdghf}.}. This expression implies that at high temperature, the thermal noise has a white spectrum to make the kernel function local in time, and that the coupling, in the form of $\gamma_{i}$, is sufficiently weak such that any contribution that depends on the frequency cutoff is ignorable. 

In this limit we arrive at a much simpler set of differential equations
\begin{align}
	\frac{d}{dt}\bm{\sigma}_{\chi_i\chi_i}&= \frac{2}{m} \,\bm{\sigma}_{\chi_ip_i}\,, \label{covar-ohmic-1st} \\ 
	\frac{d }{dt}\bm{\sigma}_{\chi_i\chi_j}&= \frac{1}{m} \,\bm{\sigma}_{p_i\chi_j}+ \frac{1}{m} \,\bm{\sigma}_{\chi_ip_j}\,, \\
	\frac{d }{dt}\bm{\sigma}_{\chi_ip_i} &= \frac{1}{m}\bm{\sigma}_{p_ip_i} -2\gamma_{i}\bm{\sigma}_{\chi_ip_i} - m\omega^{2}\bm{\sigma}_{\chi_i\chi_i}-m\sigma(t)\bm{\sigma}_{\chi_i\chi_j}\,,\\
	\frac{d }{dt}\bm{\sigma}_{\chi_ip_j}&  = \frac{1}{m}\bm{\sigma}_{p_ip_j}- 2\gamma_{j}\bm{\sigma}_{\chi_ip_j}-m\omega^2\bm{\sigma}_{\chi_i\chi_j} -m\sigma(t)\bm{\sigma}_{\chi_i\chi_i}\,,  \\
	\frac{d }{dt}\bm{\sigma}_{p_ip_j}&= - 2(\gamma_{i} + \gamma_{j})\bm{\sigma}_{p_ip_j}- m\omega^2\bigl[\bm{\sigma}_{p_i\chi_j}+\bm{\sigma}_{\chi_ip_j}\bigr]-m \sigma(t)\bigl[\bm{\sigma}_{\chi_ip_i}+\bm{\sigma}_{\chi_jp_j}\bigr]\,, \\
	\frac{d }{dt}\bm{\sigma}_{p_ip_i}&= \frac{4m\gamma_{i}}{\beta_{i}^{\textsc{b}}} - 4\gamma_{i} \bm{\sigma}_{p_ip_i} - 2m\omega^2\bm{\sigma}_{\chi_ip_i} -m\sigma(t)\bm{\sigma}_{p_i\chi_j}\,,\label{covar-ohmic-last}
\end{align}
with $j=3-i$ and $i=1$, 2.
 \begin{figure}
 	\centering
	\scalebox{1.5}{\includegraphics[width=0.45\textwidth]{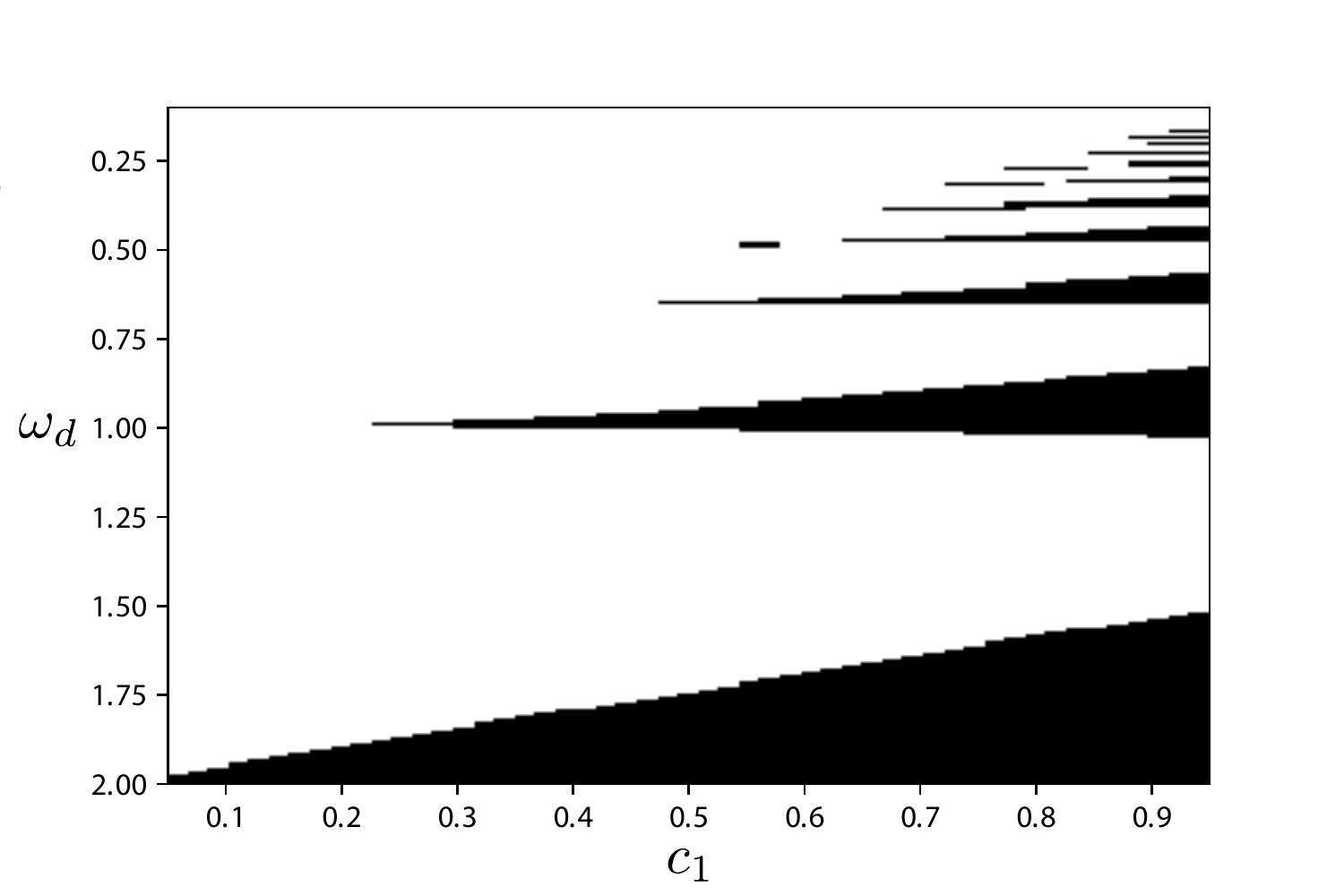}}
	\caption{\label{fig:stability} Phase diagram for stability in terms of driving frequency $\omega_{d}=\frac{2\pi}{\tau_{d}}$ and coupling parameter $c_{1}$ when $\sigma(t)$ takes the form $\sigma(t)=c_{0}+c_{1}\,\cos\omega_{d}t$. Black regions in the diagram are unstable while the white regions are stable. The constant parameters are $m = 1$, $\omega = 1$, $\gamma_{1}=\gamma_{2}= 0.005$, and $c_0 = 0$.}
 \end{figure} 
Before proceeding to investigating parametrically driven open system dynamics, let us comment on the stability of the quantum Langevin equation \eqref{lange-sf}. Following Ref.~\cite{mathieu-stbl}, we will define a matrix $\bm{C}$ that describes the movement in the phase space over one driving period $\tau_d $ from the initial phase space position $\bm{Z}(0)$. We first rewrite Eqs.~\eqref{q-1st-mom-eqn} and \eqref{p-1st-mom-eqn} in terms of this phase space variable $\bm{Z}(t)$,
\begin{align}
	\langle\dot{\bm{Z}}(t)\rangle&=\bm{W}(t)\cdot \langle\bm{Z}(t)\rangle\,,&\bm{W}(t)&= \begin{prmatrix}  
 0 &0 &m^{-1} &0 \\ 0 &0 &0 &m^{-1} \\ -m\omega^2 &-m\sigma(t) &-2\gamma_{1} &0 \\ -m\sigma(t) &-m\omega^2 &0 &+2\gamma_{2} \end{prmatrix}\,.
\end{align}
This implies that we may introduce the matrix $\bm{\mathfrak{D}}(t)$, in a similar fashion as the fundamental solutions $\bm{D}_{i}(t)$ in the configuration space, that maps the initial state $\bm{Z}(0)$ to the current state $\bm{Z}(t)$ by $\langle\bm{Z}(t)\rangle=\bm{\mathfrak{D}}(t)\cdot\langle\bm{Z}(0)\rangle$. The matrix $\bm{D}$ satisfies $\dot{\bm{\mathfrak{D}}}(t)=\bm{W}(t)\cdot\bm{\mathfrak{D}}(t)$, and is explicitly related to the fundamental solution matrices $\bm{D}_{i}$ by
\begin{equation}
	\bm{\mathfrak{D}} = \begin{pmatrix} \bm{D}_1 &\bm{D}_2 \\ \dot{\bm{D}}_1  &\dot{\bm{D}}_2      \end{pmatrix}\,.
\end{equation}
Then, the matrix $\bm{C}$ is defined as $\bm{C} = \bm{\mathfrak{D}}(\tau_d)$ and the stability is defined by the requirement that all eigenvalues of $\bm{C}$ should have the modulus less than one~\cite{mathieu-stbl,mathieu-stbl2}. Fig.~\ref{fig:stability} shows a phase diagram of the parameter space $(c_{1},\omega_{d})$. The black shade represents the dynamically unstable regime, while the white area denotes the stable regime.

\subsection{Entanglement dynamics}\label{entng-ohmic}

We now study the evolution of the entanglement of the coupled oscillators, in contact with their own private high temperature baths. We use logarithmic negativity as the entanglement measure~\cite{part-tr,Plenio}. This is achieved by numerically solving the coupled evolution equations of the covariance matrix elements, \eqref{covar-ohmic-1st}--\eqref{covar-ohmic-last}. 
The logarithmic negativity is defined as $E_{\mathcal{N}} = \max\{0,-\ln(2\lambda_{<}^{\textsc{pt}})\}$, where $\lambda_{<}^{\textsc{pt}}$ is the smaller symplectic eigenvalue of the partially transposed covariance matrix $\bm{\sigma}^{\textsc{pt}}$. When $E_{\mathcal{N}}>0$, the oscillators are entangled.

The symplectic eigenvalues of $\bm{\sigma}^{\textsc{pt}}$ can be calculated most easily from the eigenvalues of the matrix $i\,\Sigma\cdot\bm{\sigma}^{\textsc{pt}}$ with 
\begin{equation}
	\Sigma = \begin{pmatrix} 0 &0 &+1 &0\\ 0 &0 &0 &+1 \\ -1 &0 &0 &0\\ 0 &-1 &0 &0 \end{pmatrix}\,,
\end{equation}
for our choice of phase space vector $\bm{Z}$. Thus the symplectic eigenvalues come in pairs, $(\pm\lambda_{>}^{\textsc{pt}},\pm\lambda_{<}^{\textsc{pt}})$, so  we set $\lambda_{>}^{\textsc{pt}} > \lambda_{<}^{\textsc{pt}} > 0$ for definiteness.

\section{Unstable vs stable dynamics: Instability a necessary condition}\label{numer-ohmic}

\subsection{Unstable dynamics}

\begin{figure}[h!]
 	\centering
	\scalebox{1.5}{\includegraphics[width=0.45\textwidth]{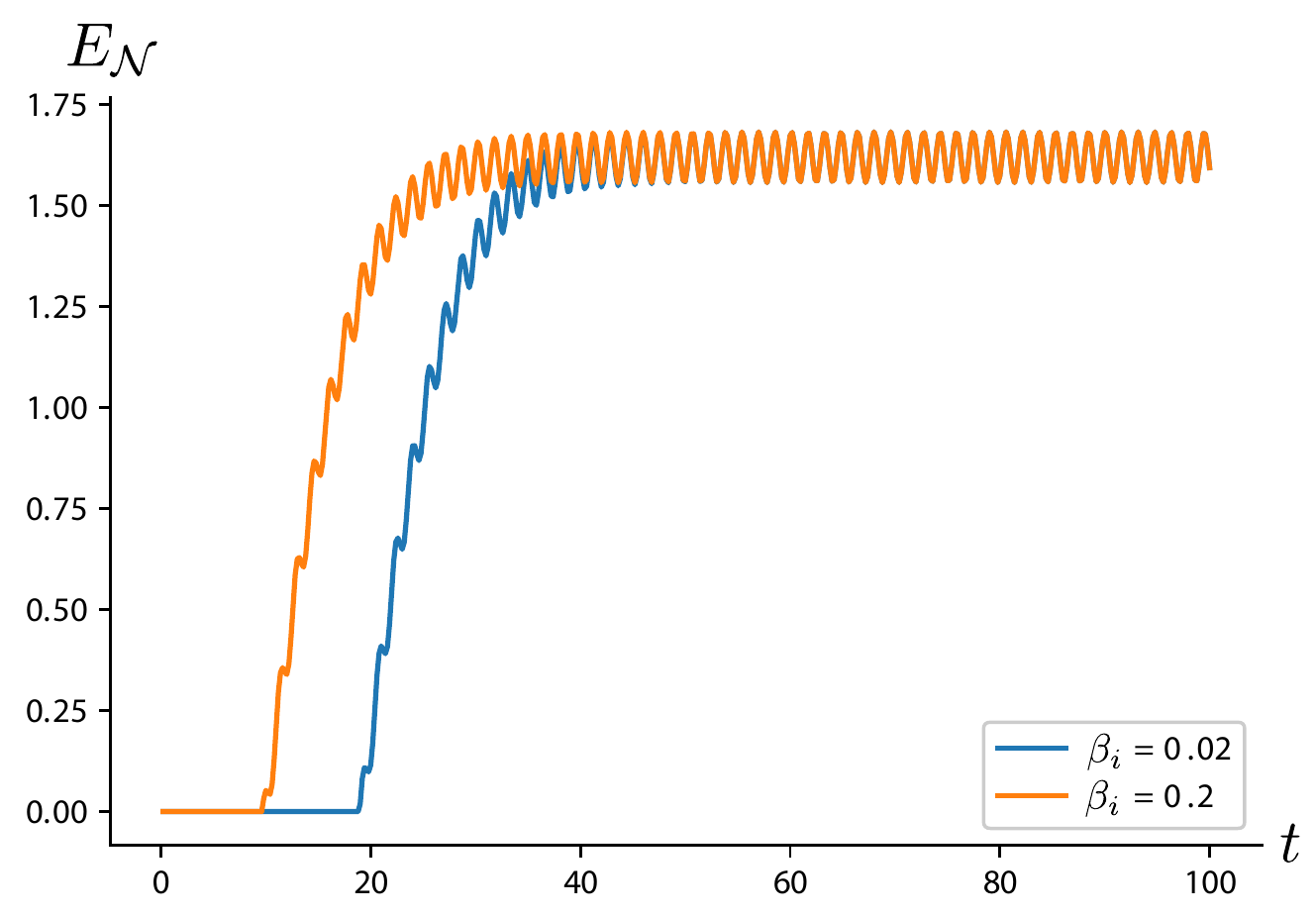}}
	\caption{\label{fig:logneg-galve} Time evolution of the logarithmic negativity when we choose the parameters by $m = 1$, $ \omega = 1$, $\gamma_{1}=\gamma_{2}= 0.0025$, $\beta_{1}^{\textsc{b}}=\beta_{2}^{\textsc{b}}= 0.2$, $\sigma(t) = c_0 + c_1\cos(\omega_d t)$ with $c_0 = 0$, $c_1 = 0.5$, $\omega_d = 1.996$. The parameter $\beta_i$ in the plot legend denotes the initial inverse temperature of the oscillators.}
 \end{figure}

We consider a bath that has an Ohmic spectral density function with infinite cutoff frequency,  thus a Markovian bath.  We choose the same set of parameters as used in Fig.~2(b) of Ref.~\cite{galve-prl} to first reproduce how the logarithmic negativity changes with time\footnote{Note  we use natural logarithm in the definition of logarithmic negativity whereas Ref.~\cite{galve-prl}  defines it with base 2. In addition, our definition of the decay coefficients $\gamma$  differs from that in Ref.~\cite{galve-prl} by a factor of 2.} in Fig.~\ref{fig:logneg-galve}. Our primary purpose here is to make explicit the relation between hot entanglement and unstable open system dynamics.

In Fig.~\ref{fig:logneg-galve} we assume that both oscillators are initially prepared in thermal states of the same temperature $\beta_i^{-1}$. This initial oscillator temperature is high with respect to the oscillator's natural frequency $\omega$, so the thermal fluctuations completely destroy the initial entanglement between the oscillators. This can be seen from the plot that $E_{\mathcal{N}}$ is essentially zero at the beginning.  We also let the bath temperature $(\beta_{i}^{\textsc{b}}){^{-1}}$ set in the high temperature regime. In the conventional setting considered in~\cite{HH15PRD},  at late times the entanglement will not survive when $\beta^{\textsc{b}}_{i}<\mathcal{O}(\omega^{-1})$. However, as have shown by Ref.~\cite{galve-prl} and seen in Fig.~\ref{fig:logneg-galve}, the entanglement is still sustained at late times in this high temperature setting for the parametrically coupled oscillators.

Fig.~\ref{fig:logneg-galve} furthermore confirms the conclusion of Ref.~\cite{galve-prl} that logarithmic negativity at late times does not depend on the initial state of the oscillators. This kind of phenomenon usually occurs when the dynamics has a steady state. There, the decaying behavior of the fundamental solution in \eqref{chi} and \eqref{p} renders any dependence on the initial state negligibly small at late times. Thus it may be surprising to see this phenomenon even when the dynamics is unstable. The instability of the system can be inferred from the observation that the parameters $\omega_{d}$, $c_{0}$, and $c_{1}$ used in Fig.~\ref{fig:logneg-galve} fall inside the dark shade in Fig.~\ref{fig:stability}. When the system is unstable, the fundamental solutions in \eqref{chi} and \eqref{p} will grow indefinitely, so it is natural to expect that the contributions from the initial state to become more and more significant. But here lies the subtlety: for example, the covariance matrix elements of the system can be divided into an active component and a passive component. The former depends on the initial conditions,  and describes the system's intrinsic behavior even when the interaction with the environment is turned off. The latter is generated by the bath after the evolution starts, and is independent of the initial condition. Thus, if the passive component of {the quantities of interest} dominates over the active component, then at late times this quantity, though still growing with time, can barely depend on its initial condition.

Now let us inspect a little more the symplectic eigenvalue $\lambda_{<}^{\textsc{pt}}$ of the partially transposed covariance matrix. It can be expressed by two symplectic invariants $\Delta(\bm{\sigma}^{\textsc{pt}})$ and $\det \bm{\sigma}$~\cite{PPTSimon,Illuminati04},
\begin{align}
	(\lambda_{<}^{\textsc{pt}})^{2} = \frac{1}{2}\left\{ \Delta(\bm{\sigma}^{\textsc{pt}}) - \sqrt{\Delta^2(\bm{\sigma}^{\textsc{pt}}) - 4\det \bm{\sigma}}  \right\}\,, \label{sympeigen-exp}
\end{align}
with $\Delta(\bm{\sigma}^{\textsc{pt}}) =  I_{1}+I_{2}-2I_{3}$, and
\begin{align}
    I_1 &=\bm{\sigma}_{\chi_1\chi_1}\bm{\sigma}_{p_1p_1} -\frac{1}{4}\bm{\sigma}_{\chi_1p_1}^2\,,  &I_2 &=\bm{\sigma}_{\chi_2\chi_2}\bm{\sigma}_{p_2p_2} -\frac{1}{4}\bm{\sigma}_{\chi_2p_2}^2\,,\notag \\
    I_3 &=\bm{\sigma}_{\chi_1\chi_2}\bm{\sigma}_{p_1p_2}-\bm{\sigma}_{\chi_1p_2}\bm{\sigma}_{\chi_2p_1}\,.  \label{sympeigen-exp-det}
\end{align}
\begin{figure}
  	\begin{subfigure}{0.49\textwidth}
     	\includegraphics[width=\textwidth]{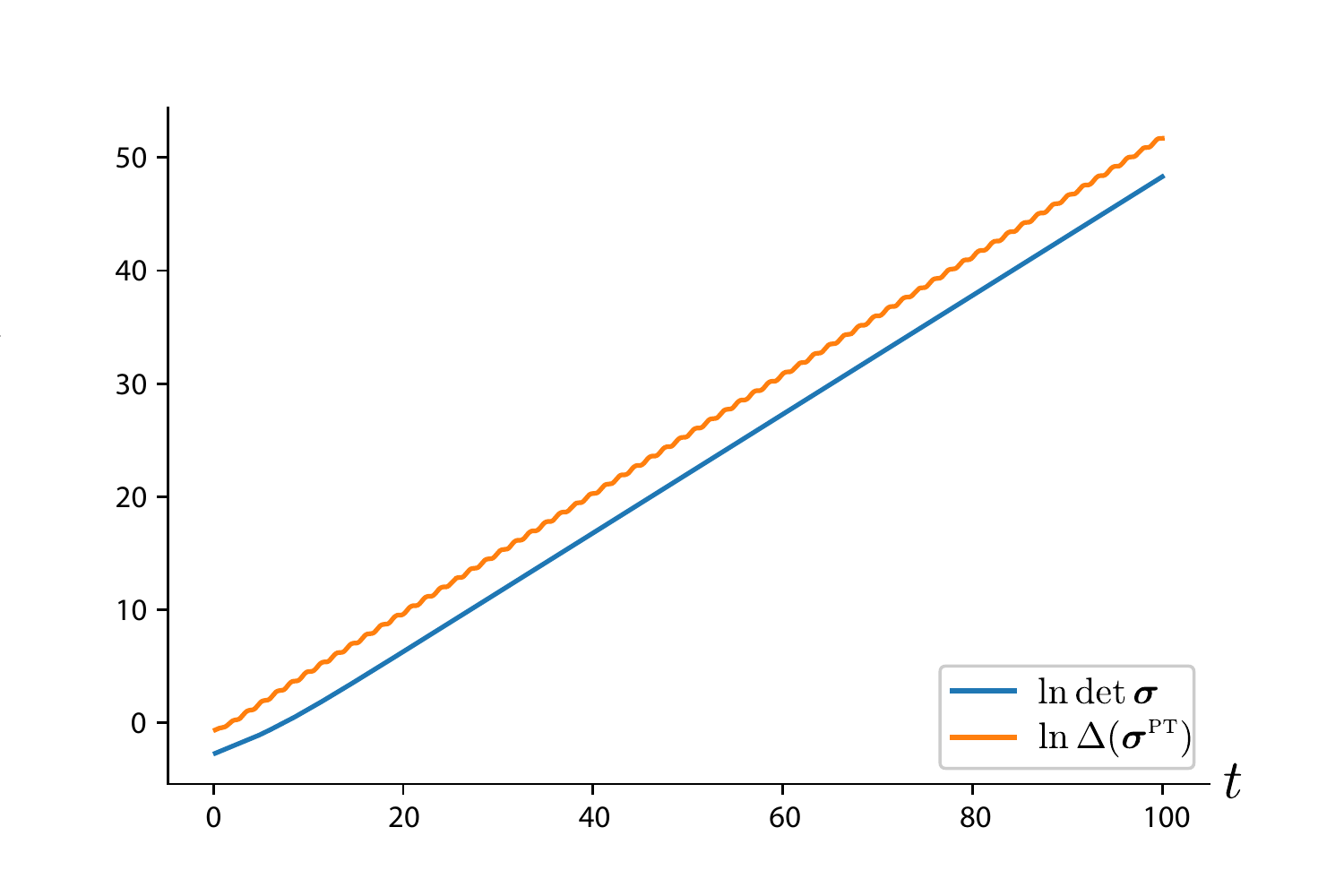}     
     	\caption{with private baths}
		\label{fig:galve-sympeigen-terms-wbath}
	\end{subfigure}
 	\hfill
	\begin{subfigure}{0.49\textwidth}
     	\includegraphics[width=\textwidth]{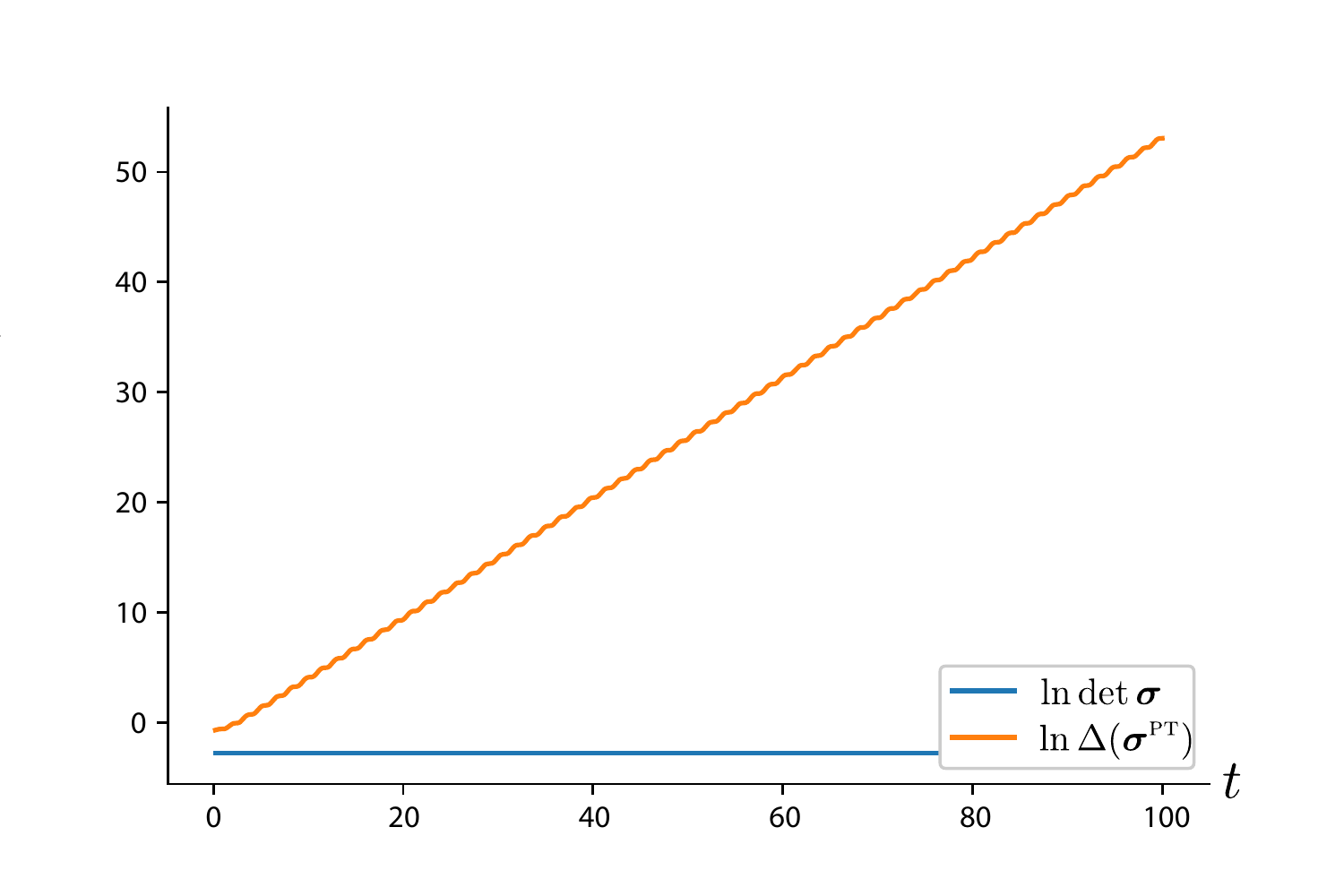}
     	\caption{without private baths}
		\label{fig:galve-sympeigen-terms-wobath}
	\end{subfigure}
	\caption{Evolution of $\Delta(\bm{\sigma}^{\textsc{pt}})$ and $\det\bm{\sigma}$ of $(\lambda_{<}^{\textsc{pt}})^{2}$ given in Eq.~\eqref{sympeigen-exp}. The same parameters as in  Fig.~\ref{fig:logneg-galve}  are used except that $\gamma = 0.005$ in (a). In {contrast, in} (b), the coupled oscillators form a closed system without contact with the thermal baths.}
\label{fig:galve-sympeigen-terms}
\end{figure}
In Fig.~\ref{fig:galve-sympeigen-terms}, we show the time evolution of $\Delta(\bm{\sigma}^{\textsc{pt}})$ and $\det\bm{\sigma}$ that constitute of $(\lambda_{<}^{\textsc{pt}})^{2}$ according to Eq.~\eqref{sympeigen-exp}. We observe that both $\Delta(\bm{\sigma}^{\textsc{pt}})$ and $\det \bm{\sigma} $ in Eq.~\eqref{sympeigen-exp} grow exponentially with time. Intriguingly the difference between $\ln\Delta(\bm{\sigma}^{\textsc{pt}})$ and $\ln\det \bm{\sigma}$ seems to remain constant with time when $t$ is sufficiently large. From this difference we can infer that 
\begin{align}\label{E:dfhgvd}
	\ln\Delta(\bm{\sigma}^{\textsc{pt}})&>\ln\det \bm{\sigma}\,,&&\Rightarrow&2\ln\Delta(\bm{\sigma}^{\textsc{pt}})&>\ln\Delta(\bm{\sigma}^{\textsc{pt}})>\ln\det\bm{\sigma}\,,\notag\\
 &&&\Rightarrow&\Delta^{2}(\bm{\sigma}^{\textsc{pt}})&\gg\det\bm{\sigma}\,.
\end{align}
This is an important criterion. Since entanglement occurs when $(\lambda_{<}^{\textsc{pt}})^{2}<1/4$, it implies that $\Delta(\bm{\sigma}^{\textsc{pt}})$ is barely greater than $\sqrt{\Delta^{2}(\bm{\sigma}^{\textsc{pt}})-4\det \bm{\sigma}}$. This is possible only when $\Delta^{2}(\bm{\sigma}^{\textsc{pt}})\gg4\det \bm{\sigma}$. In this case the square of the symplectic eigenvalue $\lambda_{<}^{\textsc{pt}}$ can be approximately given by
\begin{equation}
    (\lambda_{<}^{\textsc{pt}})^2\simeq \frac{\det\bm{\sigma}}{\Delta(\bm{\sigma}^{\textsc{pt}})}\,.
\end{equation}
Thus entanglement exists when $\det\bm{\sigma}<\Delta(\bm{\sigma}^{\textsc{pt}})/4$. Later we will give another example in which the dynamics is unstable but $\Delta^{2}(\bm{\sigma}^{\textsc{pt}})\gtrsim4\det \bm{\sigma}$. In that case entanglement is not realizable at high temperatures.

The observation $\ln\Delta(\bm{\sigma}^{\textsc{pt}})-\ln\det \bm{\sigma}\simeq\text{const}>0$ implies that for sufficiently large time, $(\lambda_{<}^{\textsc{pt}})^{2}$ will be a constant smaller than unity. On the other hand, the fact that entanglement can be sustained when $(\lambda_{<}^{\textsc{pt}})^{2}<1/4$ in turn tells us that the difference between $\ln\Delta(\bm{\sigma}^{\textsc{pt}})$ and $\ln\det \bm{\sigma}$ needs to be greater than $\ln4\sim1.39$. Apparently this requirement is well satisfied in Fig.~\ref{fig:galve-sympeigen-terms-wbath}, not to mention Fig.~\ref{fig:galve-sympeigen-terms-wobath}. The small ripples in the curve of $\ln\Delta(\bm{\sigma}^{\textsc{pt}})$ does not affect the above arguments; they are manifested in the oscillatory behavior of $E_{\mathcal{N}}$ in Fig.~\ref{fig:logneg-galve}.

Fig.~\ref{fig:galve-sympeigen-terms-wobath} shows that the determinant of the covariance matrix remains constant for the closed system case because the evolution is unitary when the bath is absent. It is fixed by our choice of the initial states of the oscillator. As $\Delta(\bm{\sigma}^{\textsc{pt}})$ keeps growing under the influence of parametric coupling, the symplectic eigenvalue $\lambda_{<}^{\textsc{pt}}$ in Eq. \eqref{sympeigen-exp} will approach zero, equivalent to an infinite logarithmic negativity. That is, entanglement is well preserved. Comparing Figs.~\ref{fig:galve-sympeigen-terms-wbath} with \ref{fig:galve-sympeigen-terms-wobath}, we may conclude that the growth of $\det \bm{\sigma}$ clearly results from the intervention of the baths. However, at this point it is not yet {fully} clear why 1) $\Delta(\bm{\sigma}^{\textsc{pt}})$ barely depends on the baths, 2) $\ln\Delta(\bm{\sigma}^{\textsc{pt}})-\ln\det \bm{\sigma}$ is a constant (apart from small ripples), and 3) the late-time entanglement is independent of the initial states when the oscillators are parametrically coupled. {We conjecture that 3) may result from the possibility that the contributions which depend on the initial condition play a subdominant role.} This scenario is particularly likely when the bath temperature is high {since the contributions from the bath can be amplified by the high temperature}.

\begin{figure}
  	\begin{subfigure}{0.49\textwidth}
     	\includegraphics[width=\textwidth]{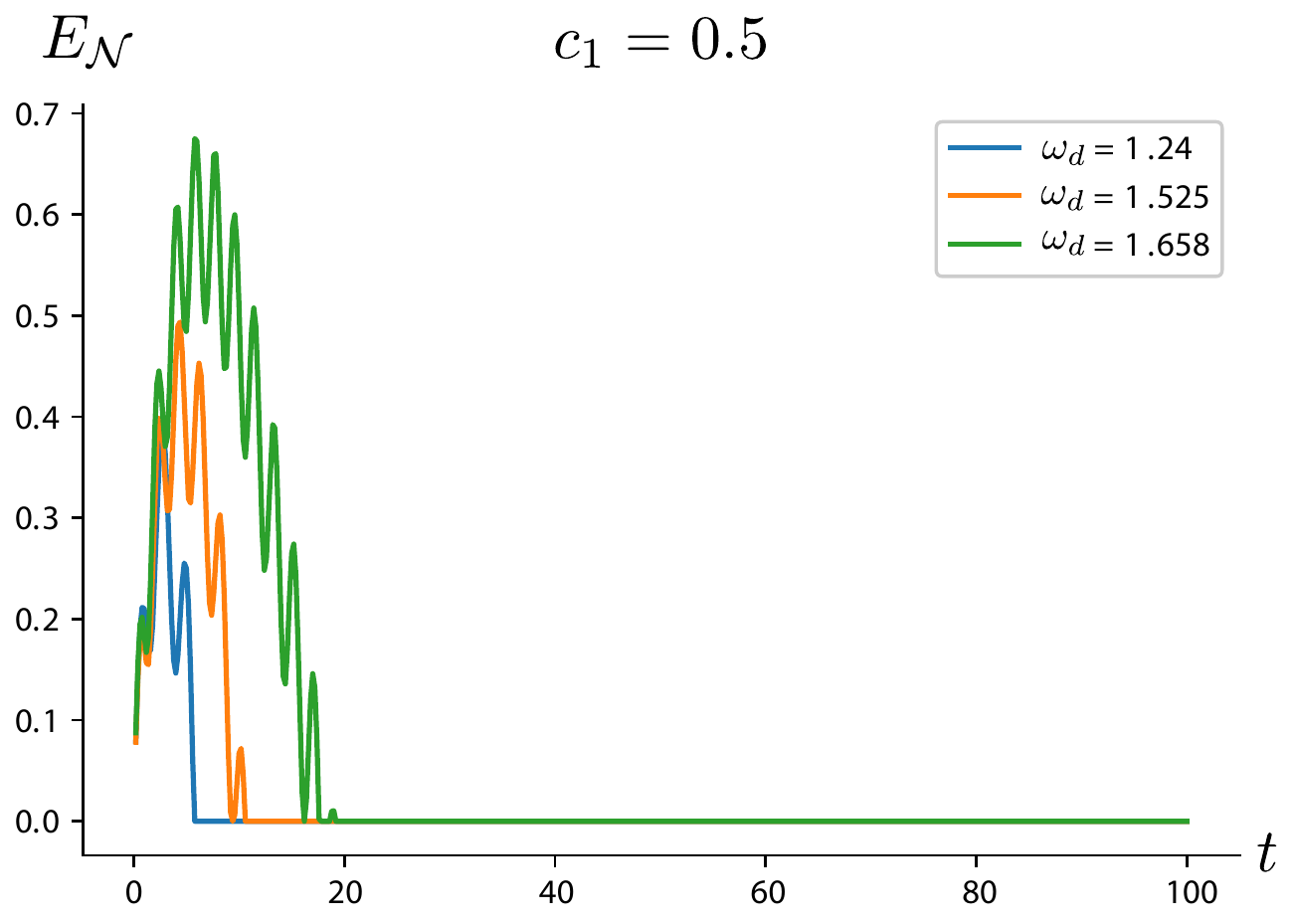}     
     	\caption{}
		\label{fig:stable-wd} 
	\end{subfigure}
 	\hfill
	\begin{subfigure}{0.49\textwidth}
     	\includegraphics[width=\textwidth]{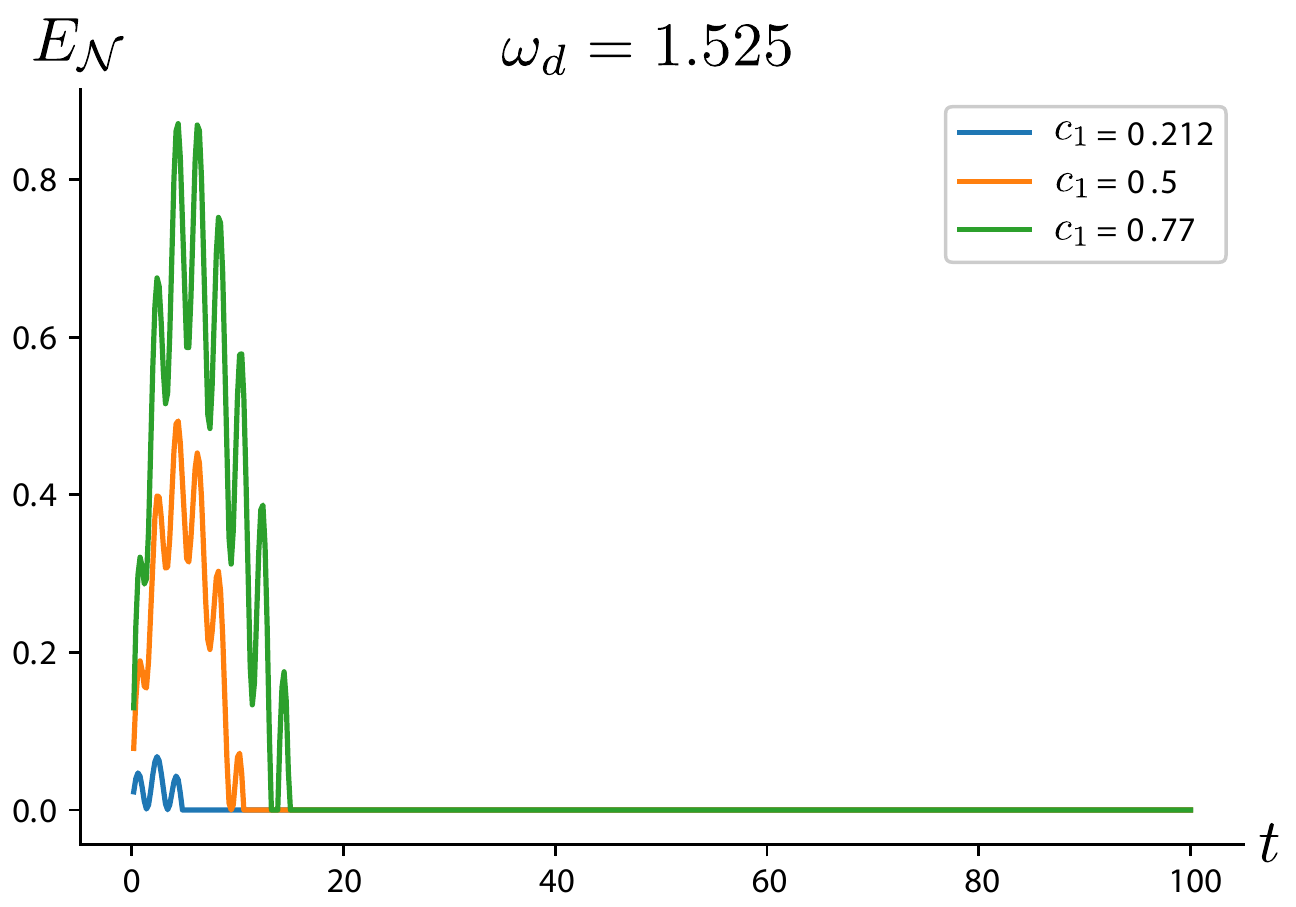}
     	\caption{}
		\label{fig:stable-str} 
	\end{subfigure}
	\caption{Time evolution of the logarithmic negativity at high bath temperature in the stable regime for different (a) driving frequencies and (b) coupling strengths. The parameters shared in both cases are $m = 1$,  $\omega = 1$, $\gamma_{1}=\gamma_{2}= 0.005$, $\beta^{\textsc{b}}_{1}=\beta^{\textsc{b}}_{2}= 0.2$, $\beta_i = 2\times 10^4$, $c_0 = 0$, but in (a) $c_1 = 0.5$ and in (b) $\omega_d = 1.525$.}
	\label{Fi:ebgtsd}
\end{figure}

\subsection{Stable dynamics}
Next we turn to stable dynamics. In this case, the contributions from the initial conditions in \eqref{chi} and \eqref{p} will be exponentially small at late times due to the decaying behavior of $\bm{D}_{i}(s,0)$. However, the periodicity property of $\bm{D}_{i}(s,s')$ in \eqref{E:dkjgbfkse} implies that the contributions from the quantum fluctuations of the private baths tend to be periodic in time as well. Thus in this regime, the covariance matrix elements will evolve to have finite, but periodic behavior. This is in strong contrast to the situation that the system is not parametrically driven. In the latter case, the observables of the system tends to rest on constant values {or constant rates} at late times in the stationary state. Therefore, in the current case when the system is parametrically driven by a periodic external agent and has stable dynamics, we understand only in an average sense that the system reaches a \textit{stationary} state. That is, the system's observables will appear to be constants only after we average them over the driving period at late times. With this dynamic feature in mind, we now examine the entanglement in the stable regime.

\begin{figure}
    \begin{subfigure}{0.49\textwidth}
     	\includegraphics[width=\textwidth]{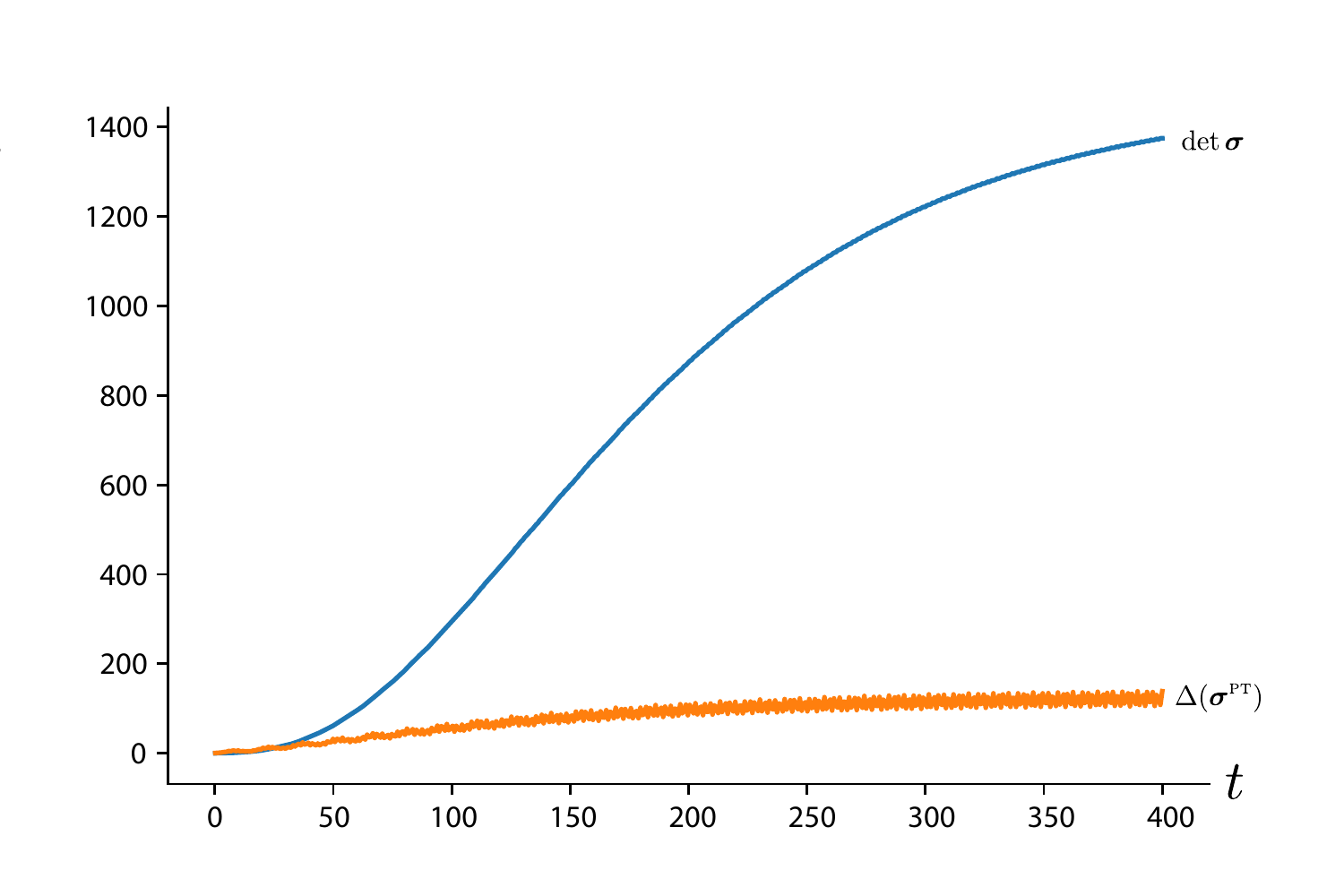}     
     	\caption{with the baths}
		\label{fig:galve-sympeigen-stable-terms-wbath} 
	\end{subfigure}
 	\hfill
	\begin{subfigure}{0.49\textwidth}
     	\includegraphics[width=\textwidth]{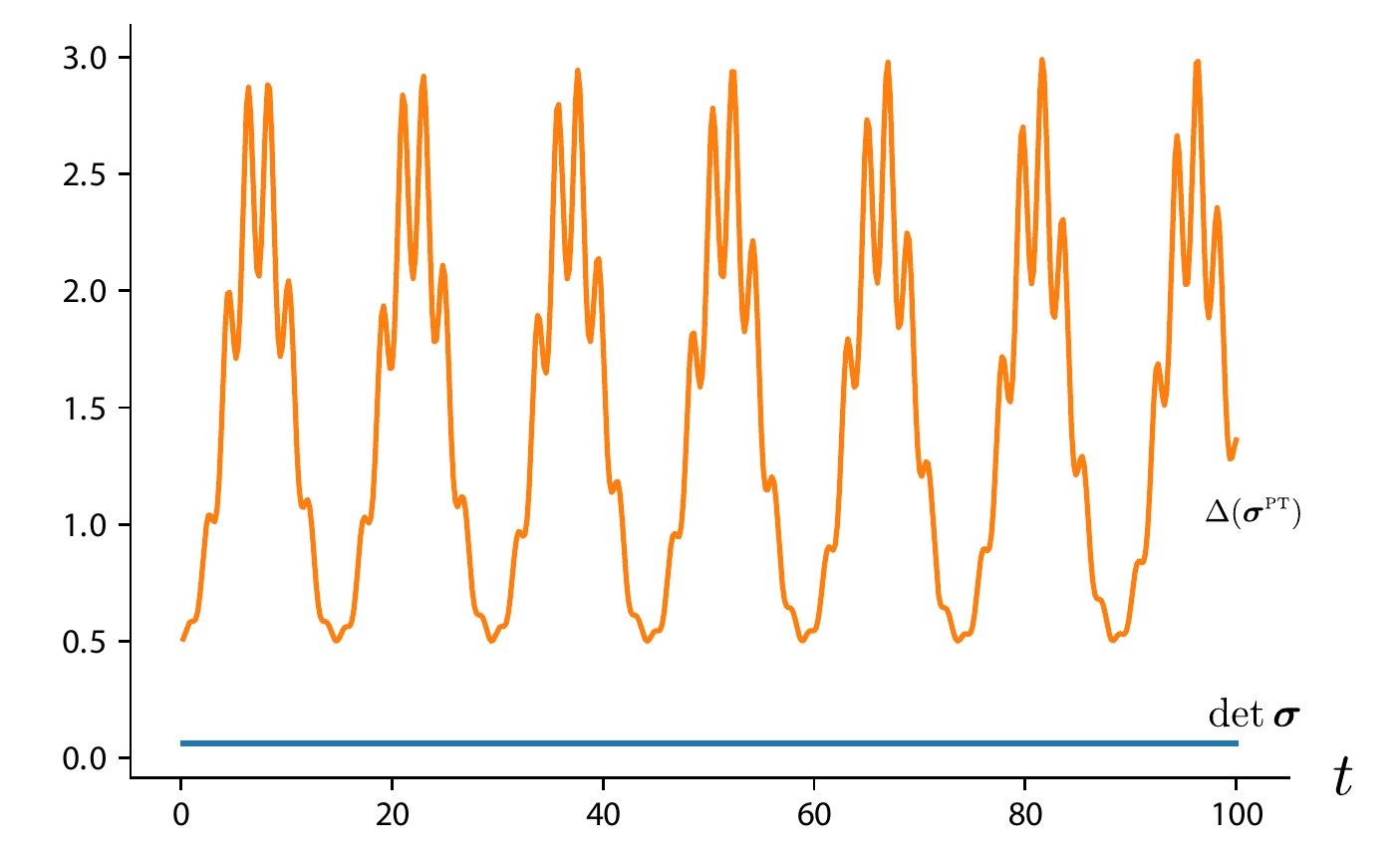}
     	\caption{without the baths}
		\label{fig:galve-sympeigen-stable-terms-wobath} 
	\end{subfigure}

    \begin{subfigure}{0.49\textwidth}
     	\includegraphics[width=\textwidth]{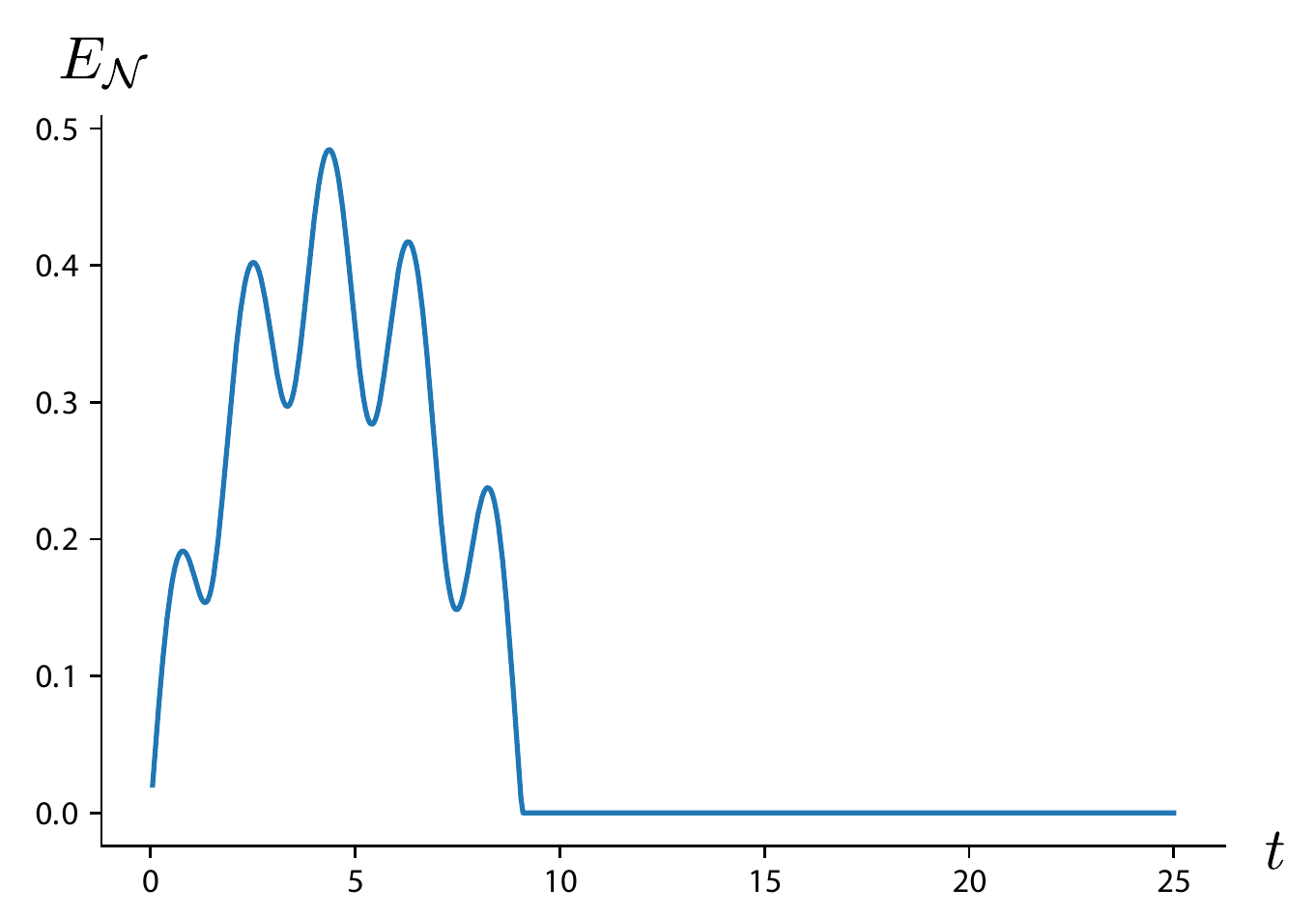}     
     	\caption{with the baths}
		\label{fig:logneg-fig5a} 
	\end{subfigure}
 	\hfill
	\begin{subfigure}{0.49\textwidth}
     	\includegraphics[width=\textwidth]{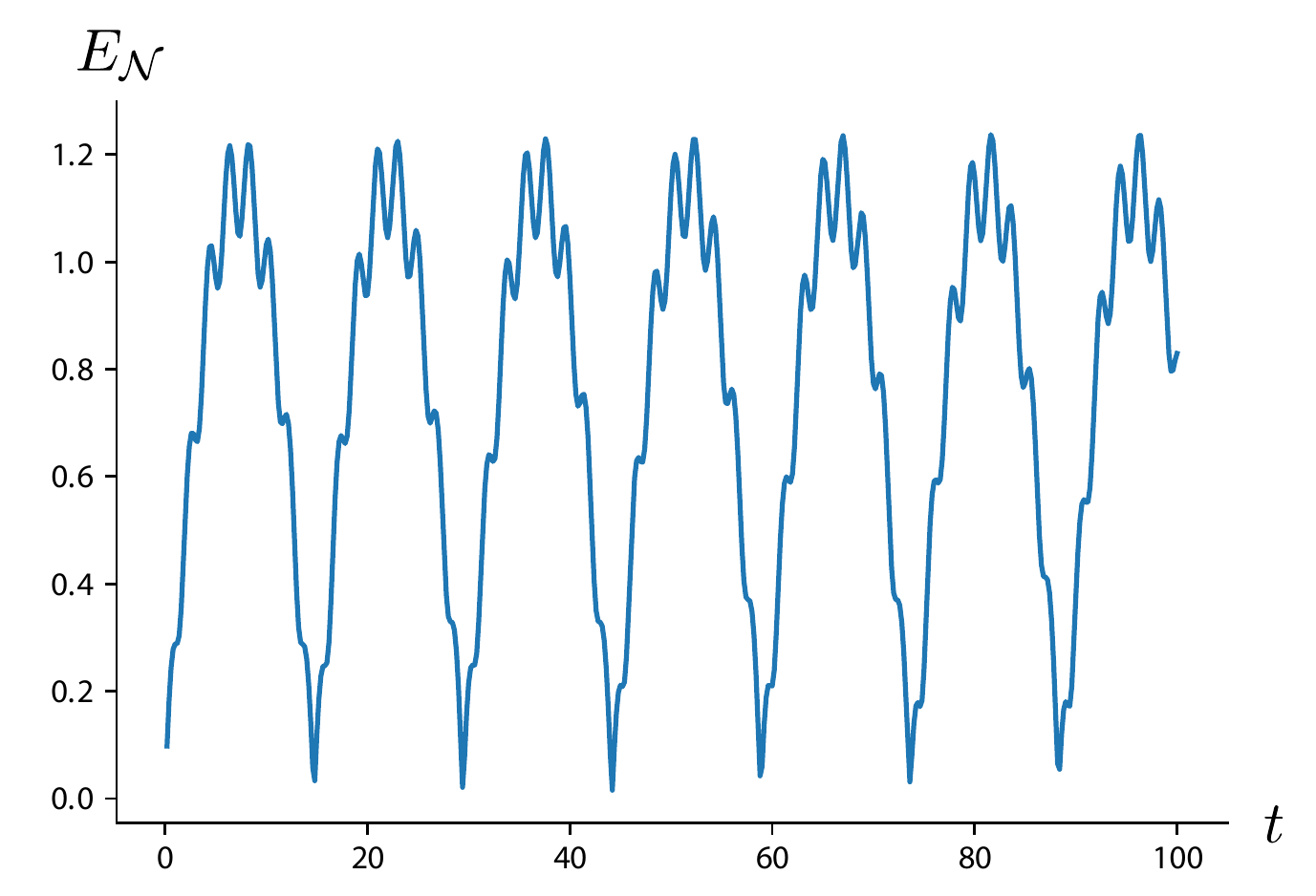}
     	\caption{without the baths}
		\label{fig:galve-det-wobath-stable} 
	\end{subfigure}
	\caption{First row:~evolution of $\Delta(\bm{\sigma}^{\textsc{pt}})(t)$ and $\det \bm{\sigma}(t)$ where the parameters $m=1$, $\omega=1$, $\gamma_{i} = 0.005$, $\beta_i^{\vphantom{\textsc{b}}}=2\times10^4$, $\beta_i^{\textsc{b}}=0.2$, $\omega_d = 1.5$, $c_0=0$ and $c_1=0.5$ are chosen for stable evolution of the system. In (a) the parametrically coupled oscillators are in contact with their private thermal bath, but in (b) the contact is removed. Second row:~evolution of the corresponding logarithmic negativity. In (c) the entanglement drops to zero rather quickly, compared with the $\gamma_i^{-1}$.}
	\label{fig:galve-sympeigen-stable-terms}
\end{figure}

Fig.~\ref{fig:stable-wd} shows the time evolution of logarithmic negativity at high bath temperature in the stable regime for different driving frequencies and Fig.~\ref{fig:stable-str} shows the time evolution of logarithmic negativity in the stable regime for different coupling strengths.

We see that the entanglement measure drops to zero very rapidly, compared to the time scale $\gamma_{i}^{-1}\sim200$, which is the typical relaxation time when the system is not parametrically driven. Before the entanglement measure vanishes, it seems that we will have greater transient entanglement when the parameters $(c_{1},\omega_{d})$ are located closer to the boundary between the stable and the unstable regimes in Fig.~\ref{fig:stability}. We further note that after the sudden death~\cite{YuEberly} of quantum entanglement, it never revives at late times in the high bath temperature $(T^{(\textsc{b})}_{1,2}\sim5\omega)$ case. This is consistent with the findings in the non-parametric driving case in~\cite{HH15PLB,Anders,AndWin,HH15JHEP}. The evolution of $\Delta(\bm{\sigma}^{\textsc{pt}})$ and $\det \bm{\sigma} $ in the expression, Eq.~(\ref{sympeigen-exp}), of $\lambda_{<}^{2}$ is plotted in Fig.~\ref{fig:galve-sympeigen-stable-terms}. We observe that  $\Delta(\bm{\sigma}^{\textsc{pt}})$ keeps oscillating around an average value for both open and closed system cases, {where in the open-system case, shown in Fig.~\ref{fig:logneg-fig5a}, the state of the coupled system rapidly becomes separable, not long after the evolution starts}. Thus, following the discussion around the criterion \eqref{E:dfhgvd}, we can draw the conclusion that {in the open system setting of Fig.~\ref{fig:galve-sympeigen-stable-terms}, since} $\Delta(\bm{\sigma}^{\textsc{pt}})\gg1$ and $\Delta^{2}(\bm{\sigma}^{\textsc{pt}})\gtrsim4 \det \bm{\sigma}$, the symplectic eigenvalues of the partially transposed covariance matrix satisfy $\lambda_{>}\gtrsim\lambda_{<}\gg1$ and entanglement { may not survive} at late times.

\begin{figure}
  	\begin{subfigure}{0.49\textwidth}
     	\includegraphics[width=\textwidth]{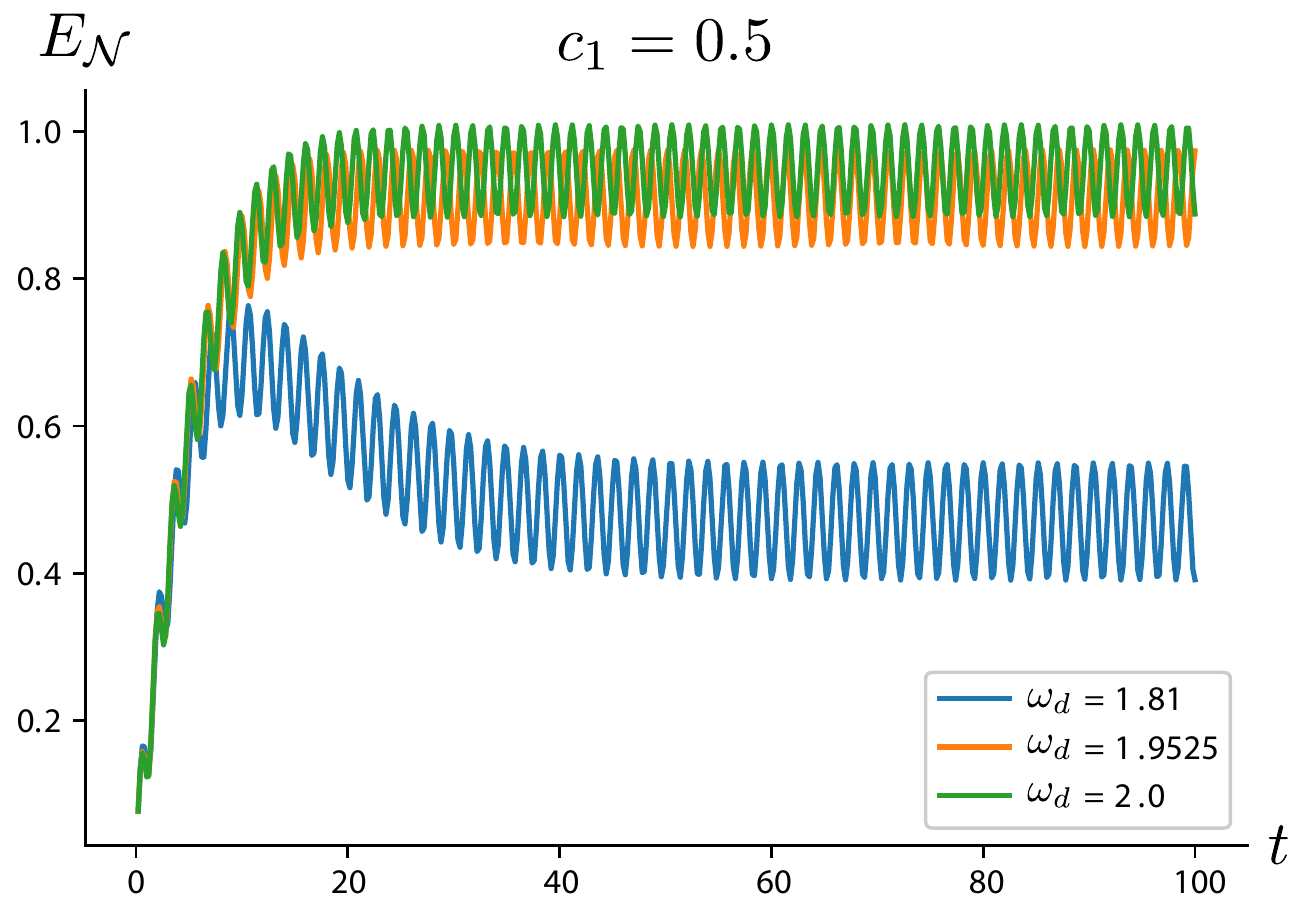}     
     	\caption{}
		\label{fig:unstable-wd} 
	\end{subfigure}
 	\hfill
	\begin{subfigure}{0.49\textwidth}
     	\includegraphics[width=\textwidth]{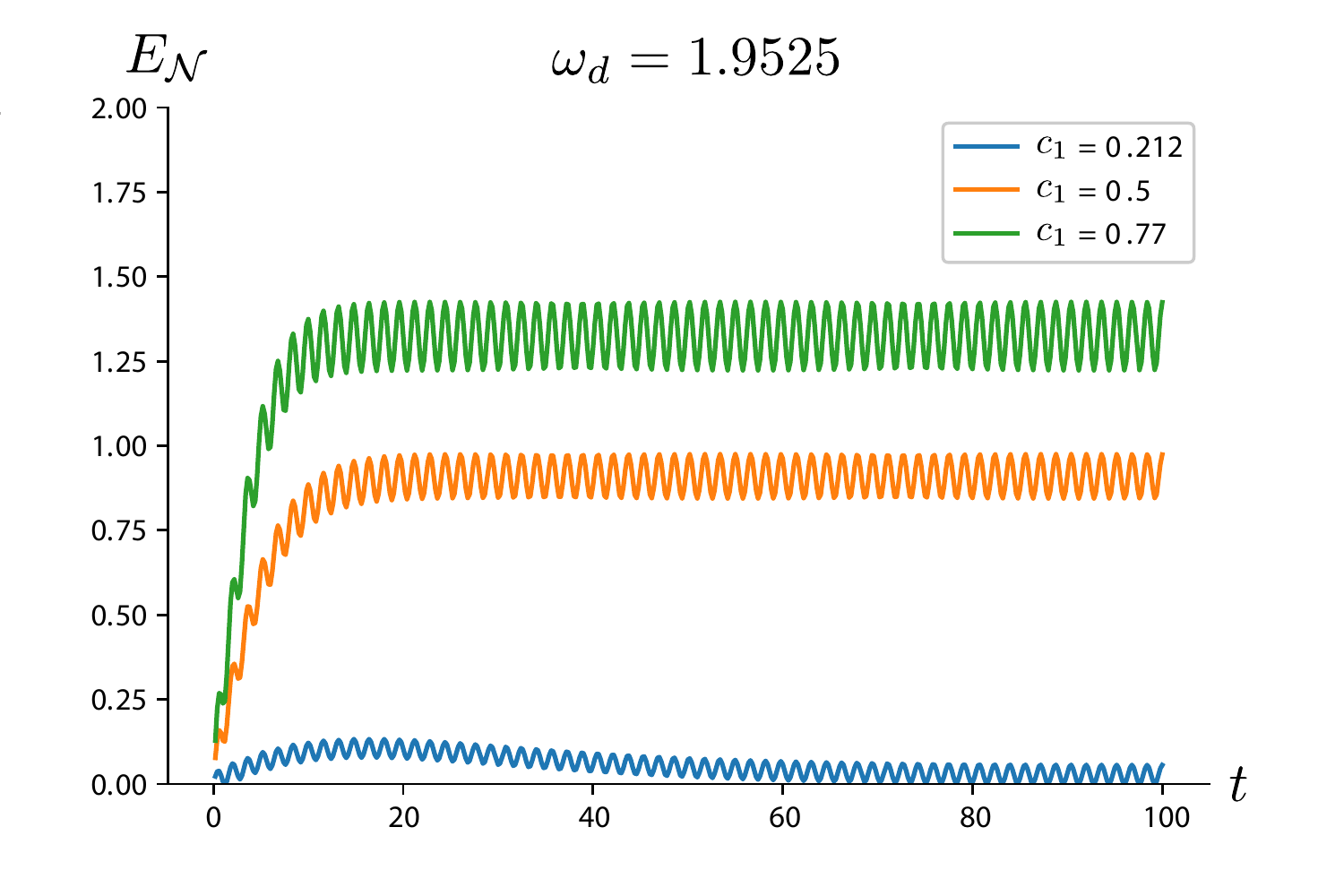}
     	\caption{}
		\label{fig:unstable-str} 
	\end{subfigure}
	\caption{Time evolution of the logarithmic negativity in the unstable regime for  $m = 1$, $\omega = 1$, $\gamma_{1}=\gamma_{2}= 0.005$, $\beta^{\textsc{b}}_{1}=\beta^{\textsc{b}}_{2}= 0.2$, $\beta_i = 2\times 10^4$, and $c_0 = 0$. In (a) we choose $c_1 = 0.5$ and in (b) $\omega_d = 1.9525$.}
	\label{fig:galve-sympeigen-stable-terms2}
\end{figure}

For comparison with Fig.~\ref{Fi:ebgtsd}, we show the evolution of logarithmic negativity, when the motion of coupled oscillator is unstable, for different driving frequencies in Fig.~\ref{fig:unstable-wd} and different $c_{1}$ in Fig.~\ref{fig:unstable-str}. In addition, we also observe a similar phenomenon, seen in Fig.~\ref{Fi:ebgtsd}, that when the parameter pair $(c_1,\omega_{d})$ is more deeply located inside the dark regime, the sustained entanglement seems stronger because the corresponding negativity is greater~\cite{Eisert98,Virmani}. Thus it seems that dynamical instability is a necessary ingredient to sustain a non-zero logarithmic negativity, allowing quantum entanglement to last over long times.

\section{Instability is not a sufficient condition}\label{S:eoujhbdf}

Here we give an example to show that dynamical instability, though necessary, is not sufficient to ensure high-temperature entanglement. Following that,  we propose an alternative but equivalent description for entanglement, in terms of squeezing and thermal fluctuations.

\subsection{Coupled amplifying harmonic oscillators}

 Let us consider the coupled amplifying harmonic oscillators, each interacting with its individual bath, following the equations of motion given by
\begin{align}
	\ddot{\chi}_{1}(t)-2\mathsf{g}\,\dot{\chi}_{1}(t)+\omega^{2}\chi_{1}(t)+c_{0}\,\chi_{2}(t)&=\frac{e}{m}\,\xi_{1}(t)\,,\label{E:eotugbsd1}\\
	\ddot{\chi}_{2}(t)-2\mathsf{g}\,\dot{\chi}_{2}(t)+\omega^{2}\chi_{1}(2)+c_{0}\,\chi_{1}(t)&=\frac{e}{m}\,\xi_{2}(t)\,.\label{E:eotugbsd2}
\end{align}
Since this phenomenological model is used to illustrate a point, we may leave out questions about the mechanism of amplification. Here, `amplifying' or `anti-damping'  means the opposite of `damping': the system oscillator's dynamical equation has a damping term with a sign opposite to that in a normal oscillator. We assume that both oscillators in our system have the same mass $m$, physical oscillating frequency $\omega$, amplification constant $\mathsf{g}$, and the same coupling strength $e$ with their individual private bath. The strength, denoted by $c_{0}$, of inter-oscillator coupling is assumed to be time-independent.  The noise force $\xi_{i}$ accounts for the quantum fluctuations of the private bath $i$, which initially is assumed to be in a thermal state of temperature $\beta^{-1}$. The corresponding dissipative backreaction of the private bath is assumed to be overwhelmed by the unspecified amplification mechanism, so that their overall effect is given by $-2\mathsf{g}\,\dot{\chi}_{i}(t)$ in the equations of motion. In principle $\mathsf{g}$ is not necessarily equal to $\gamma=e^{2}/(8\pi m)$, but we assume that it takes on this value for the moment. We still suppose the baths satisfy the standard fluctuation-dissipation relation associated with their initial state.

This choice of parameters allows us to easily decouple the equations of motion into
\begin{align}
	\ddot{\chi}_{\pm}(t)-2\mathsf{g}\,\dot{\chi}_{\pm}(t)+\omega^{2}_{\pm}\chi_{\pm}(t)&=\frac{e}{m}\,\xi_{\pm}(t)\,,\label{E:fgbkdgye}
\end{align}
where $\omega_{\pm}^{2}=\omega^{2}\pm\sigma$ by the normal modes
\begin{equation}
	\chi_{\pm}(t)=\frac{1}{\sqrt{2}}\,\chi_{1}(t)\pm\frac{1}{\sqrt{2}}\,\chi_{2}(t)\,.
\end{equation}
Since the Langevin equations \eqref{E:fgbkdgye} are homogeneous in time, the fundamental solutions are given by
\begin{align}
	d_{1}^{(\pm)}(t)&=e^{\mathsf{g}t}\Bigl[\cos\Omega_{\pm}t-\frac{\mathsf{g}}{\Omega_{\pm}}\,\sin\Omega_{\pm}t\Bigr]\,,&d_{2}^{(\pm)}(t)&=e^{\mathsf{g}t}\,\frac{1}{\Omega_{\pm}}\sin\Omega_{\pm}t\,,
\end{align}
with $\Omega_{\pm}^{2}=\omega_{\pm}^{2}-\mathsf{g}^{2}=\omega^{2}-\mathsf{g}^{2}\pm\sigma$. They grow exponentially due to the amplification effect. In this case $D_{2}^{(\pm)}(t,s)=d_{2}^{(\pm)}(t-s)$. The  behavior  of the covariance matrix elements may not be completely controlled by the bath at late time because the contributions from the initial conditions can be significant due to runaway dynamics. However, their relative dominance depends on the bath temperature.

\begin{figure}
\centering
    \scalebox{0.35}{\includegraphics{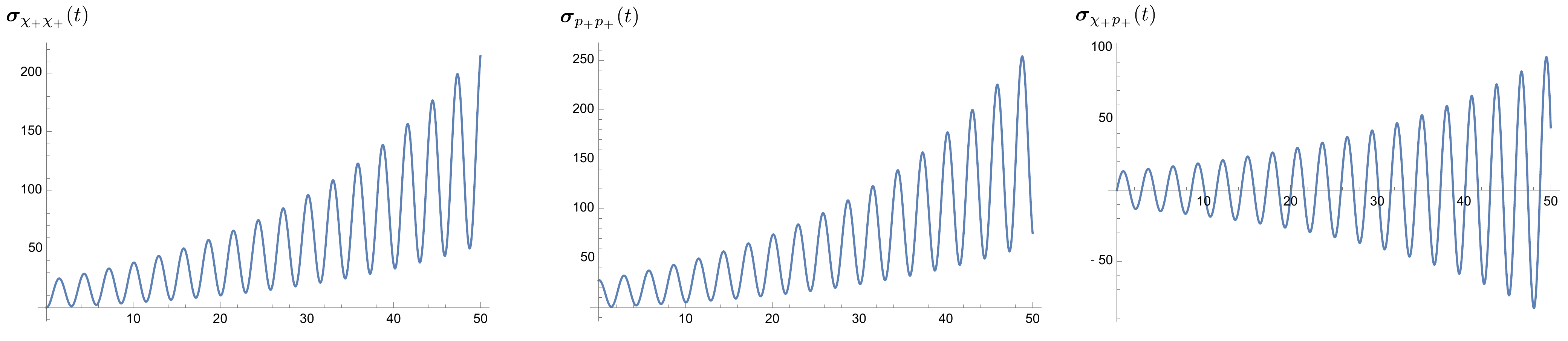}}
    \caption{The time evolution of the covariance matrix elements $\bm{\sigma}_{\chi_{+}\chi_{+}}(t)$, $\bm{\sigma}_{p_{+}p_{+}}(t)$ and $\bm{\sigma}_{\chi_{+}p_{+}}(t)$ for the coupled amplifying oscillators in the NESS configuration. We choose the parameters $m=1$, $\mathsf{g}=\gamma=0.02$, $\sigma=0.2$, and $\eta=2$. The temperature of two private bath is $\beta^{-1}=10$, which belongs to the conventional high temperature regime. All the parameters and the evolution time are normalized to $\omega$ or $\omega^{-1}$.}\label{Fi:AO_ppMode}
\end{figure}
Let us consider the time evolution of the covariant matrix elements. Suppose that the initial state of the amplifying oscillators is a two-mode squeezed vacuum state, so both oscillators are initially entangled, Thus in the normal modes, $\bm{Z}_{\pm}=(\chi_{+}, p_{+}, \chi_{-}, p_{-})^{T}$, the covariance matrix element has the form
\begin{align*}
	\bm{\sigma}_{\pm}(0)&=\begin{pmatrix}\sigma_{\chi_{+}\chi_{+}}(0) &0 &0 &0 \\[4pt]0 &\sigma_{p_{+}p_{+}}(0) &0 &0\\[4pt]0 &0 &\sigma_{\chi_{-}\chi_{-}}(0) &0\\[4pt] 0 &0 &0 &\sigma_{p_{-}p_{-}}(0) \end{pmatrix}\,,
\end{align*}
with 
\begin{align*}
	\sigma_{\chi_{+}\chi_{+}}(0)&=\frac{1}{2m\omega}e^{-2\eta}\,,&\sigma_{p_{+}p_{+}}(0)&=\frac{m\omega}{2}e^{+2\eta}\,,\\
	\sigma_{\chi_{-}\chi_{-}}(0)&=\frac{1}{2m\omega}e^{+2\eta}\,,&\sigma_{p_{+}p_{+}}(0)&=\frac{m\omega}{2}e^{-2\eta}\,.
\end{align*}
Here for example, $\sigma_{\chi_{+}\chi_{+}}$ represents
\begin{equation}
	\sigma_{\chi_{+}\chi_{+}}=\frac{1}{2}\langle\{\chi_{+},\,\chi_{+}\}\rangle
\end{equation}
and $\eta$ is the {initial} squeeze parameter. Since both modes are decoupled, the covariance matrix is always in block form:
\begin{align}\label{E:pworihfbd}   
    \bm{\sigma}_{\pm}(t)&=\begin{pmatrix}\bm{A}_+(t) &\bm{0}\\\bm{0} &\bm{A}_-(t)\end{pmatrix}\,,&\bm{A}_{\pm}&=\begin{pmatrix}
        \bm{\sigma}_{\chi_{\pm}\chi_{\pm}} &\bm{\sigma}_{\chi_{\pm}p_{\pm}}\\\bm{\sigma}_{\chi_{\pm}p_{\pm}} &\bm{\sigma}_{p_{\pm}p_{\pm}}
    \end{pmatrix}\,,
\end{align}
with elements like $\bm{\sigma}_{\chi_{\pm}\chi_{\pm}}(t)$ given by
\begin{align}
    \bm{\sigma}_{\chi_{\pm}\chi_{\pm}}(t)&=d_1^{(\pm)}(t)\,\bm{\sigma}_{\chi_{\pm}\chi_{\pm}}(0)+\frac{1}{m^2}\,d_2^{(\pm)}(t)\,\bm{\sigma}_{p_{\pm}p_{\pm}}(0)\notag\\
    &\qquad\qquad\qquad\qquad+\frac{e^2}{m^2}\int_0^t\!ds\int_0^t\!ds\;d_2^{(\pm)}(t-s)d_2^{(\pm)}(t-s')\,G_{H,0}^{(++)}(s,s')\,,
\end{align}    
where
\begin{equation}
    G_{H,0}^{(ij)}(s,s')=\frac{1}{2}\,\langle\xi_i(s)\xi_j(s')\rangle
\end{equation}
and $i$, $j=+$, $-$. The other elements in $\bm{A}_\pm$ can be constructed similarly.

\begin{figure}
\centering
    \scalebox{0.5}{\includegraphics{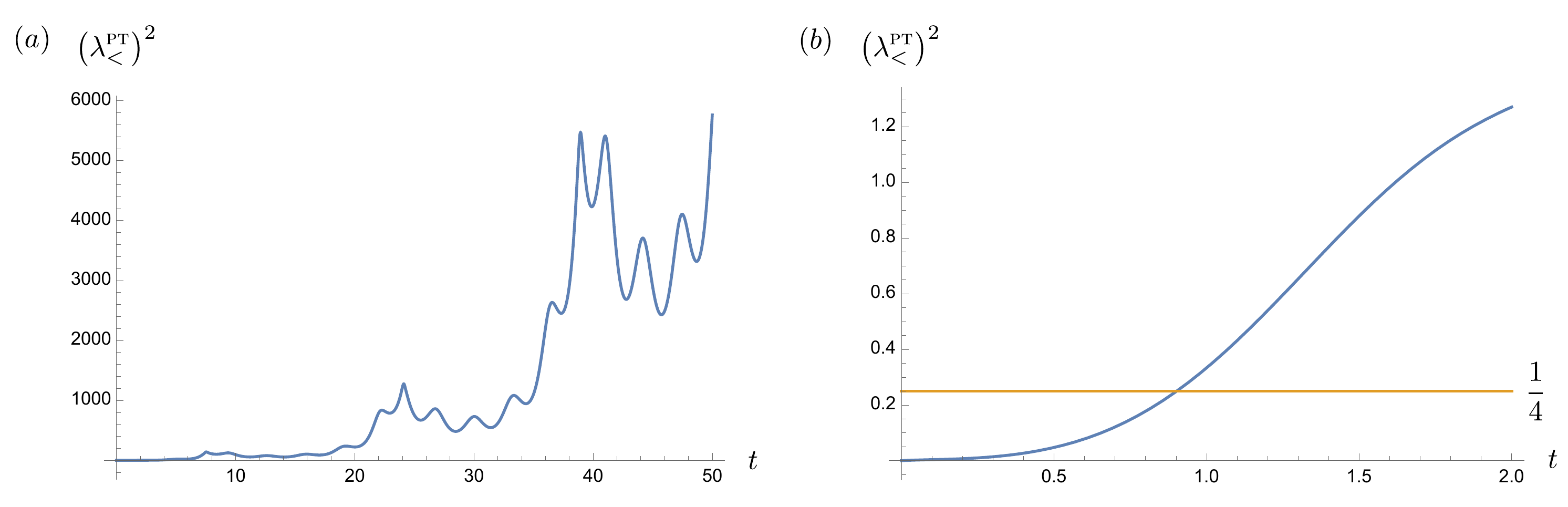}}
    \caption{The time evolution of the smaller of the squared symplectic eigenvalues of the partially transposed covariance matrix in terms of the canonical variables. Its value quickly rises past $1/4$ after roughly one cycle of motion. Afterwards, the entanglement of the partite system vanishes, and the eigenvalue seem to increase indefinitely. We choose the parameters $m=1$, $\mathsf{g}=\gamma=0.02$, $\sigma=0.2$, and $\eta=2$. The temperature of two private bath is $\beta^{-1}=10$, which belongs to the conventional high temperature regime. All the parameters and the evolution time are normalized to $\omega$ or $\omega^{-1}$.}\label{Fi:AO_sympPT}
\end{figure}

Fig.~\ref{Fi:AO_ppMode} shows the generic behavior of the covariance matrix elements with time. They are oscillatory but with increasing amplitudes, indicating runaway dynamics. When we turn to the time evolution of $(\lambda_{<}^{\textsc{pt}})^{2}$ in Fig.~\ref{Fi:AO_sympPT},  a distinct difference shows. The eigenvalue rapidly grows out of bound. When we zoom into the early time regime, we observe that the initial entanglement, $(\lambda_{<}^{\textsc{pt}})^{2}<1/4$, is destroyed right after about one cycle of motion. Thus it is barely sustained, in strong contrast to the unstable parametric driven case in~\cite{galve-prl}.
 
The results shown in Fig.~\ref{Fi:AO_sympPT} hint that
\begin{align}\label{E:gfkdbsert}
	\lambda^{\textsc{pt}}_{>}&\gtrsim\lambda^{\textsc{pt}}_{<}\,,&&\text{but}&\lambda^{\textsc{pt}}_{>}\lambda^{\textsc{pt}}_{<}&\gg1\,.
\end{align}
According to \eqref{sympeigen-exp}, they in turns imply
\begin{align}
	\Delta(\bm{\sigma}^{\textsc{pt}})&\gg1\,, &\det\bm{\sigma}&\gg1\,,&&\text{but} &\Delta^{2}(\bm{\sigma}^{\textsc{pt}})\gtrsim4\det\bm{\sigma}\,,
\end{align}
such that $\Delta(\bm{\sigma}^{\textsc{pt}})\gg\sqrt{\Delta^{2}(\bm{\sigma}^{\textsc{pt}})-4\det\bm{\sigma}}$, since
\begin{align}
	\bigl(\lambda^{\textsc{pt}}_{>}\bigr)^{2}\bigl(\lambda^{\textsc{pt}}_{<}\bigr)^{2}=\det\bm{\sigma}\,.
\end{align}
Comparing with the unstable parametrically driven oscillator case, even this cursory analysis shows that in order to have sustained entanglement, we need to have a configuration that keeps $\det\bm{\sigma}$ in check such that $\Delta^{2}(\bm{\sigma}^{\textsc{pt}})\gg4\det\bm{\sigma}$. It is exactly the case found in \eqref{E:dfhgvd} for the unstable, parametrically driven case. This ensures the contribution from the thermal fluctuations is subdominant. It seems to tell that cross correlation between the oscillators is so strong as to overcome the fluctuation effect in each oscillator.

Through this example we see that dynamical instability is a necessary but not sufficient condition for hot entanglement, it also illustrates the simple criterion in the link between dynamical instability and sustained entanglement.

\subsection{Entanglement expressed in terms of quantum optics parameters}

 Here we propose an interpretation of entanglement in terms of squeezing and thermal fluctuations parameters. Following the discussion in Appendix~\ref{S:peotr}, we notice that the covariance matrix \eqref{E:bskads} of the normal modes, which correspond to two canonical pairs that define the two mode squeezed state, has the same structure as the covariance matrix \eqref{E:pworihfbd}. It prompts the possibility to express \eqref{E:pworihfbd} at any instant by an effective two-mode squeezed state. The state corresponding to \eqref{E:pworihfbd} in fact is a product of two single-mode squeezed state, each of which rotates at different angular velocities due to different normal-mode frequencies. In this sense the two-mode squeezed state, having fewer number of independent parameters than the product of two single-mode squeezed states do, is only an effective description, requiring the effective squeeze parameter and the effective temperature to be time-dependent. Notwithstanding, the clear advantage is to describe entanglement with familiar quantities in quantum optics. We can write the symplectic eigenvalues of the partially transposed covariance matrix for the canonical pairs of the coupled oscillator as
\begin{equation}\label{E:ewtdgs}
	\lambda^{\textsc{pt}}_{\gtrless}=e^{\pm2\eta_{\textsc{eff}}}\,\bigl(\bar{n}_{\textsc{eff}}+\frac{1}{2}\bigr)\,,
\end{equation}
in analogy to \eqref{E:bskads} in terms of the effective squeeze parameter $\eta_{\textsc{eff}}$ and the effective inverse temperature $\beta_{\textsc{eff}}$, satisfying
\begin{align}
    \bar{n}_{\textsc{eff}}+\frac{1}{2}=\frac{1}{2}\,\coth\frac{\beta_{\textsc{eff}}\omega}{2}&=\sqrt{\lambda_{>}^{\textsc{pt}}\lambda_{<}^{\textsc{pt}}}\,,&\eta_{\textsc{eff}}&=\frac{1}{4}\ln\frac{\lambda^{\textsc{pt}}_{>}}{\lambda^{\textsc{pt}}_{<}}\,.
\end{align}    
\begin{figure}
\centering
    \scalebox{0.45}{\includegraphics{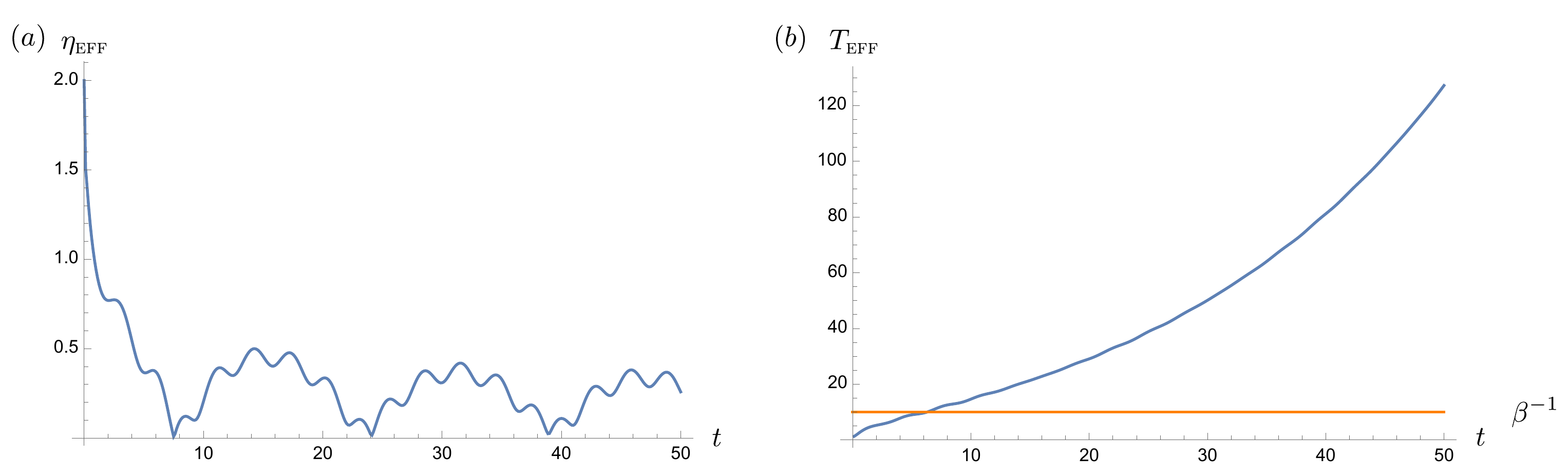}}
    \caption{The time evolution of the effective squeeze parameter $\eta_{\textsc{eff}}(t)$ and the effective temperature $T_{\textsc{eff}}(t)=\beta_{\textsc{eff}}^{-1}(t)$.  Actually, the squeeze parameter more or less decreases with time but the effective temperature grows rapidly. We choose the parameters $m=1$, $\mathsf{g}=\gamma=0.02$, $\sigma=0.2$, and $\eta=2$. The temperature of two private bath is $\beta^{-1}=10$, which belongs to the conventional high temperature regime. All the parameters and the evolution time are normalized to $\omega$ or $\omega^{-1}$.}\label{Fi:AO_effective}
\end{figure}
Here the effective temperature is determined up to a frequency factor, and we choose the physical frequency of the oscillator. If another time-independent frequency scale is chosen, it merely rescales the effective temperature but will not affect its generic feature. Both effective parameters are time-dependent, and thus are nonequilibrium in nature. They are shown in Fig.~\ref{Fi:AO_effective}. In sub-figure~\ref{Fi:AO_effective}-(a), we see the effective squeeze parameter immediately drops from its initial value, and oscillates with time, but overall it decrease with time. This could be a bad sign if we use the instability configuration to sustain entanglement. Fig.~\ref{Fi:AO_effective}-(b) shows the time evolution of the effective temperature of the system. It grows more and more rapidly with time, and zooms past the value of the initial bath temperature at roughly $t=6\omega^{-1}$. This implies that in this case instability actually induces larger and larger effective thermal fluctuations {such that they overwhelm} the effect of squeezing. This is quite different from the stable case where the contribution from thermal fluctuations approaches a constant value with time.  The compound effect of decreasing squeeze parameter and increasing thermal fluctuations makes it impossible to sustain late-time quantum entanglement of such a bipartite system in the NESS configuration.

Next we consider the energy budget in the unstable, parametrically driven case.

\section{Energy Budget operating in the unstable, parametrically driven regimes}\label{S:ebgdfhg}

Finally, in this section we scrutinize the scheme of hot entanglement proposed by Galve et al ~\cite{galve-prl} operating in the unstable regimes of parametric driven coupling from a realistic experimental feasibility perspective. We numerically study the energy flows to/from the two baths, between the coupled oscillators and calculate the work cost of driving to see if it is within reasonable experimental reach.  

\subsection{Hot entanglement can only be a transient effect from work cost considerations}

\begin{figure}
  	\begin{subfigure}{0.49\textwidth}
     	\includegraphics[width=\textwidth]{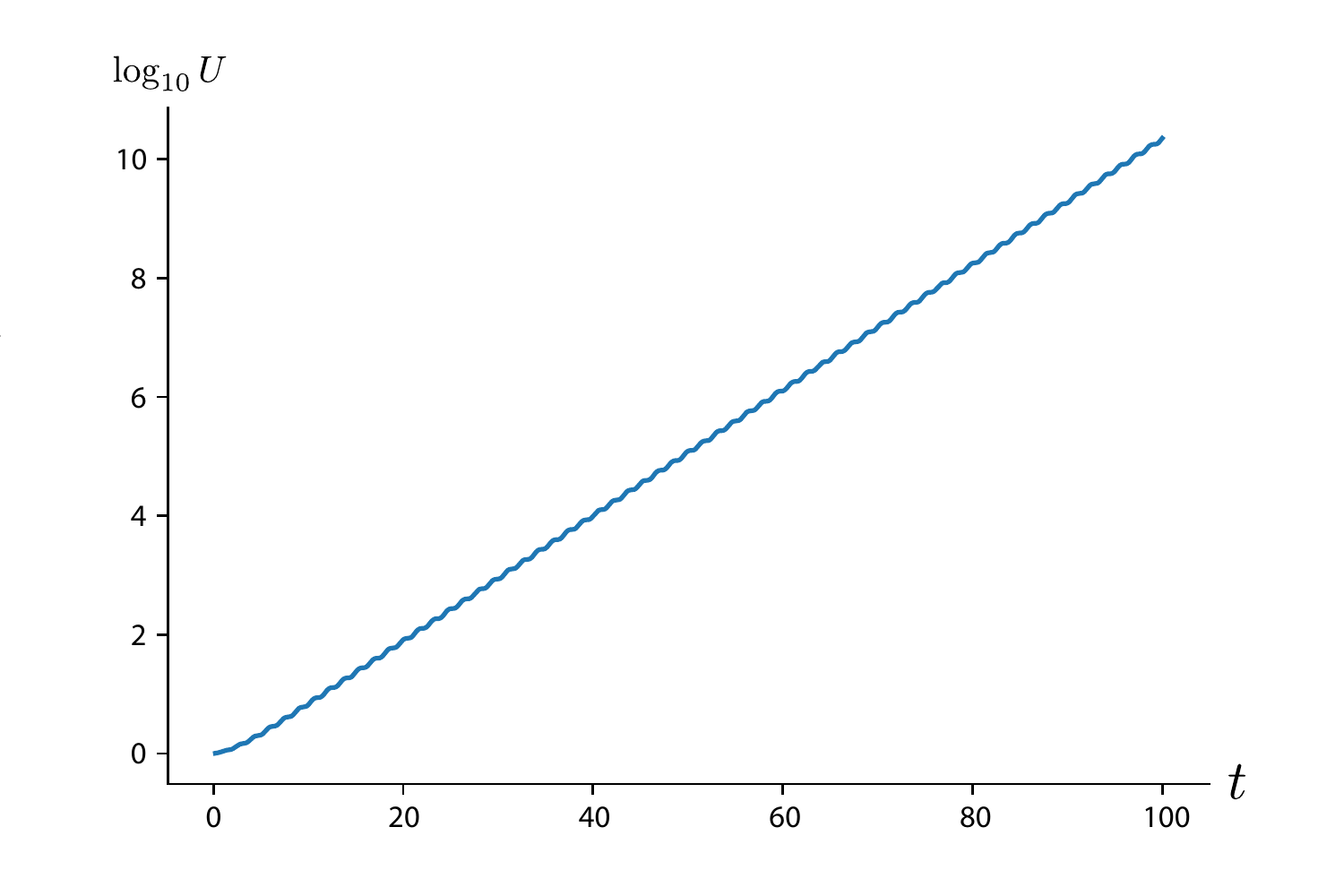}     
     	\caption{}
		\label{fig:galve-logenergy}
	\end{subfigure}
 	\hfill
	\begin{subfigure}{0.49\textwidth}
     	\includegraphics[width=\textwidth]{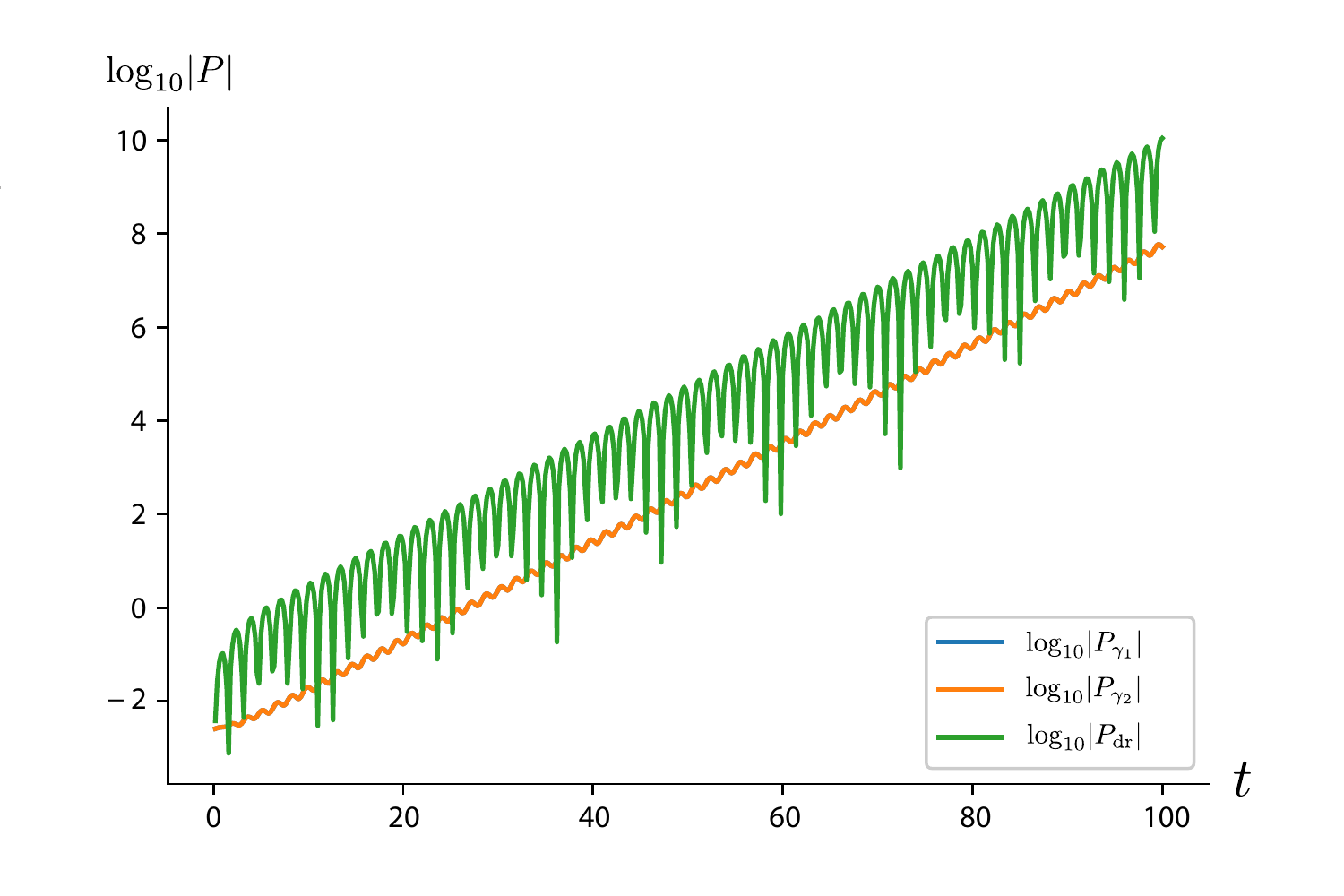}
     	\caption{}
		\label{fig:galve-logpower} 
	\end{subfigure}
	\caption{(a) Time evolution of the internal energy. (b) Time evolution of the dissipation power $\lvert P_{\gamma_{i}}\rvert$ and driving powers $P_{\text{dr}}$. Note that the dissipation power is always negative, the driving power is positive  except for a few data points  in the plot. We attribute the existence of negative values of $P_{\text{dr}}$ to numerical errors since these negative values are negligibly small. We use the same parameters aa in Fig.~\ref{fig:logneg-galve}}
	\label{fig:galve-sympeigen-stable-terms3}
\end{figure}

We first define the internal energy $U$ of the system of coupled harmonic oscillators as the expectation value of the system Hamiltonian at a given time
\begin{equation}
    U(t) =\sum_{i=1}^{2} \left( \frac{m\omega^2 \langle\chi_{i}^2\rangle}{2} + \frac{\langle p_{i}^2\rangle}{2m} \right) + m\sigma(t)\,\langle\chi_{1}\chi_{2} \rangle\, . \label{sysen}
\end{equation}
Taking the derivative of Eq.~\eqref{sysen} with respect to time and making use of Eqs.~\eqref{covar-ohmic-1st}--\eqref{covar-ohmic-last}, we obtain
\begin{align}\label{E:bkeirs}
	\frac{dU}{dt} &=\sum_{i=1}^{2} \Bigl( \frac{m\omega^2}{2} \frac{d}{dt}\langle\chi_{i}^2\rangle + \frac{1}{2m}\frac{d}{dt}\langle p_{i}^2\rangle\Bigr) + \frac{m\sigma}{2} \frac{d}{dt}\langle\{\chi_{1},\chi_{2}\}\rangle + \frac{m}{2}\frac{d\sigma}{dt}\langle\{\chi_{1},\chi_{2}\}\rangle\,.
\end{align}
On the other hand, from the Langevin equation \eqref{lange-sf}, we have
\begin{align}
    \sum_{i=1}^{2} \Bigl( \frac{m\omega^2}{2} \frac{d}{dt}\langle\chi_{i}^2\rangle + \frac{1}{2m}\frac{d}{dt}\langle p_{i}^2\rangle\Bigr) + \frac{m\sigma}{2} \frac{d}{dt}\langle\{\chi_{1},\chi_{2}\}\rangle=\sum_{i=1}^2\Bigl(\langle\xi_i\dot{\chi}_i\rangle-2m\gamma_i\langle\dot{\chi}_i^2\rangle\Bigr)\,.
\end{align}
Comparing with \eqref{E:bkeirs}, we find    
\begin{align}
	\frac{dU}{dt}&= \sum_{i=1}^{2} \Bigl(  \frac{2\gamma_{i}}{\beta^{\textsc{b}}_{i}} - \frac{2\gamma_{i}}{m} \langle p_{i}^2 \rangle \Bigr) + \frac{m}{2}\frac{d\sigma}{dt}\langle\{\chi_{1},\chi_{2}\}\rangle+\cdots\,,\label{E:kfgjbdfg}
\end{align}
where we have implemented the Caldeira-Leggett approximation and assumed the driving protocol, $\sigma(t)=c_{0}+c_{1}\,\cos\omega_{d}t$. The first term on the righthand side of \eqref{E:kfgjbdfg} describes a constant energy flow from the private baths to their respective system oscillators in the form of a white thermal noise due to our high temperature Ohmic bath assumption. The second term results from the dissipative term in Eq.~\eqref{lange-sf}, effecting an energy flow from the system oscillators to their respective baths.  The last term describes the work transfer from the external agent that performs the time-dependent driving on the system. Following the notation in Ref.~\cite{HH15AOP}, we rewrite the time derivative of the internal energy as 
\begin{equation}\label{E:kgdfge}
	\frac{dU}{dt} = \sum_{i=1}^{2} \left({P_{\xi}}_i + {P_{\gamma}}_i \right)  + P_{\text{dr}}\,.
\end{equation}
Eq.~\eqref{E:kgdfge} expresses the standard energy conservation in thermodynamics:  the work done by the external agent plus the heat from the thermal bath cause the internal energy of the system to change. With the same parameters used in Fig.~\ref{fig:logneg-galve}, we find in Figs.~\ref{fig:galve-logenergy} and \ref{fig:galve-logpower} that the internal energy $U$, dissipated power $\lvert P_{\gamma_{i}}\rvert$ and the driving power $P_{\text{dr}}$ {increase exponentially with time}. (The constant contributions $P_{\xi_{i}}$ from the bath noises are not plotted.)   The plots show that after a sufficiently long time, the system, which is composed of two parametrically coupled oscillators in the two bath configuration, draws a huge amount of energy from the external agent, resulting in the internal energy and the dissipation power to both baths grow exponentially fast. In other words, because quantum entanglement between the oscillators can be kept only in the dynamically unstable regimes, the external agent must have an exponentially large energy reservoir to {supply} the system's consumption. From a realistic experimental viewpoint this can only be supported for a short period of time, but not sustainable. We conclude that hot entanglement in this setup can only be a transient effect. 

\subsection{No fluctuation-dissipation relation at late times}

Since the energy input from the external driving agent mainly goes to the internal energy of the system and flows to the surrounding bath via the damping channel, the power delivered by the noise force from the bath plays a negligible role, even when the bath temperature is high. From this viewpoint of energy flow, the classical description seems to suffice, as if the noise term in the equation of motion \eqref{lange-sf} is absent. However, this is not true, because of the presence of quantum entanglement. The huge order of magnitude discrepancy between the dissipation power and the noise power clearly indicates it is impossible to have any steady state at late times, and thus there cannot be any fluctuation-dissipation relation for such parametrically driven systems.

\section{Conclusion}

In this paper, we probe deeper into the conditions for sustaining the entanglement between two coupled harmonic oscillators (the system) each interacting with its own heat bath, while the system is subjected to external parametric drive,  the same set up as in prior works for Markovian~\cite{galve-prl} and non-Markovian~\cite{EstPac} baths. Our results based on numerical solutions of the quantum Langevin equations show that entanglement can be sustained, but  only in the dynamically unstable regimes, and, as such, the work cost for modulating the coupling strength increases exponentially over time to make the scheme untenable. To cover a fuller ground we also investigated the possibility of sustaining entanglement in the stable regime, maximizing the effect of weak driving on the system by choosing driving frequencies slightly detuned from one of the parametric resonance frequencies. We concluded that under stable dynamics the energy cost-sustained entanglement trade-off has no particular advantage. Varying the temperature ranges we also found, with no surprise, that entanglement becomes insignificant at late times at very low temperatures. Fully considered, we are able to conclude that sustaining entanglement at high temperatures with driving protocols operating the system in the dynamically unstable regime for a long time is practically impossible due to its exponentially rising energy cost whereas operating in the stable regime requiring only constant power consumption fails to achieve hot entanglement. It remains an open challenge to investigate hot entanglement under different conditions such as making the inter-oscillator coupling or external driving nonlinear, and to try out different set ups such as sharing a common bath, and important generic systems such as two-level systems. This is an important subject which, as far as we can see, is still in its infancy awaiting more
thorough and systematic analysis. \\

\textbf{Acknowledgment} OA is supported by a research assistantship from the Maryland Center for Fundamental Physics.  J.-T. Hsiang is supported by the Ministry of Science and Technology of Taiwan, R.O.C. under Grant No.~MOST 111-2811-M-008-022. BLH appreciates the hospitality of the National Center for Theoretical Sciences of Taiwan during his visit in the finishing stage of this paper. 

\newpage

\appendix

\section{Two-mode squeezed thermal state}\label{S:peotr}

We include here, as background, for the sake of self-containment,  for two-mode squeezed thermal state, this description is adapted from~\cite{HAH22}.

Given a two-mode squeezed thermal state $\rho_{\textsc{tmsq}}^{(\beta)}$, defined by
\begin{equation}
	\rho_{\textsc{tmsq}}^{(\beta)}=\mathcal{S}_{2}^{\vphantom{\dagger}}\,\rho_{\beta}\,\mathcal{S}_{2}^{\dagger}\,,
\end{equation}
where the operator $\mathcal{S}_{2}$ is the two-mode squeeze operator $\mathcal{S}_{2}=\exp\Bigl[\zeta^{*}a_{1}^{\vphantom{\dagger}}a_{2}^{\vphantom{\dagger}}-\zeta\,a_{1}^{\dagger}a_{2}^{\dagger}\Bigr]$ with $\zeta=\eta\,e^{i\theta}$, $\eta\geq0$ and $0\leq\theta<2\pi$, we can readily find ts action on the annihilation operator $a_1$ of mode 1 given by
\begin{equation}
	\mathcal{S}_{2}^{\dagger}\,a_{1}^{\vphantom{\dagger}}\mathcal{S}_{2}^{\vphantom{\dagger}}=\cosh\eta\,a_{1}-e^{i\theta}\sinh\eta\,a_{2}^{\dagger}\,.
\end{equation}
Since the two-mode squeeze operator $\mathcal{S}_{2}$ is symmetric in $a_{1}$ and $a_{2}$, a similar result for $a_{2}$ can be found by the substitution $1\leftrightarrow2$. The annihilation operators, $a_1$ and $a_2$, of mode 1 and mode 2 satisfy the standard canonical commutation relation $[a_j^{\vphantom{\dagger}},a_k^{\dagger}]=\delta_{jk}$ with $i$, $j=1$, 2.

If we introduce the the canonical pair $(\chi_j,p_k)$ of two modes, with $[\chi_j,p_k]=i\,\delta_{jk}$, and expanded the pair by the individual annihilation and creation operators, $a_i^{\vphantom{\dagger}}$, $a_i^{\dagger}$,
\begin{align}
	\chi_{i}&=\frac{1}{\sqrt{2m\omega}}\bigl(a_{i}^{\dagger}+a_{i}^{\vphantom{\dagger}}\bigr)\,, &p_{i}&=i\sqrt{\frac{m\omega}{2}}\bigl(a_{i}^{\dagger}-a_{i}^{\vphantom{\dagger}}\bigr)\,,
\end{align}
then we find the corresponding covariance matrix elements given by
\begin{align}
	\bm{\sigma}_{\chi_1\chi_1}&=\frac{1}{m\omega}\,\Bigl[\bigl(\bar{n}_{1}+\frac{1}{2}\bigr)\,\cosh^{2}\eta+\bigl(\bar{n}_{2}+\frac{1}{2}\bigr)\,\sinh^{2}\eta\Bigr]\,,\label{E:bsgvsj1}\\
	\bm{\sigma}_{\chi_2\chi_2}&=\frac{1}{m\omega}\,\Bigl[\bigl(\bar{n}_{2}+\frac{1}{2}\bigr)\,\cosh^{2}\eta+\bigl(\bar{n}_{1}+\frac{1}{2}\bigr)\,\sinh^{2}\eta\Bigr]\,,\\
	\bm{\sigma}_{p_1p_1}&=m\omega\,\Bigl[\bigl(\bar{n}_{1}+\frac{1}{2}\bigr)\,\cosh^{2}\eta+\bigl(\bar{n}_{2}+\frac{1}{2}\bigr)\,\sinh^{2}\eta\Bigr]\,,\\
	\bm{\sigma}_{p_2p_2}&=m\omega\,\Bigl[\bigl(\bar{n}_{2}+\frac{1}{2}\bigr)\,\cosh^{2}\eta+\bigl(\bar{n}_{1}+\frac{1}{2}\bigr)\,\sinh^{2}\eta\Bigr]\,,\\
	\bm{\sigma}_{\chi_1\chi_2}&=-\frac{1}{2m\omega}\,\bigl(\bar{n}_{1}+\bar{n}_{2}+1\bigr)\,\sinh2\eta\,\cos\theta\,,\\
	\bm{\sigma}_{p_1p_2}&=+\frac{m\omega}{2}\,\bigl(\bar{n}_{1}+\bar{n}_{2}+1\bigr)\,\sinh2\eta\,\cos\theta\,,\\
	\bm{\sigma}_{\chi_1p_2}&=-\frac{1}{2}\,\bigl(\bar{n}_{1}+\bar{n}_{2}+1\bigr)\,\sinh2\eta\,\sin\theta\,,\\
	\bm{\sigma}_{\chi_2p_1}&=-\frac{1}{2}\,\bigl(\bar{n}_{1}+\bar{n}_{2}+1\bigr)\,\sinh2\eta\,\sin\theta\,,\\
	\bm{\sigma}_{\chi_1p_1}&=0\,,\\
	\bm{\sigma}_{\chi_2p_2}&=0\,.\label{E:bsgvsj10}
\end{align}
However, if we further introduce the normal-mode basis,
\begin{equation}
	\chi_{\pm}=\frac{\chi_{1}\pm\chi_{2}}{\sqrt{2}}\,,
\end{equation}
and assume the special configuration of $\bar{n}_{1}=\bar{n}_{2}=\bar{n}$, then we arrive at
\begin{align}
	\bm{\sigma}_{\chi_+\chi_+}&=\frac{1}{2m\omega}\bigl(2\bar{n}+1\bigr)\bigl(\cosh2\eta-\cos\phi\,\sinh2\eta\bigr)\,,\label{E:ikgbidehs1}\\
	\bm{\sigma}_{\chi_-\chi_-}&=\frac{1}{2m\omega}\bigl(2\bar{n}+1\bigr)\bigl(\cosh2\eta+\cos\phi\,\sinh2\eta\bigr)\,,\\
	\bm{\sigma}_{p_+p_+}&=\frac{m\omega}{2}\bigl(2\bar{n}+1\bigr)\bigl(\cosh2\eta+\cos\phi\,\sinh2\eta\bigr)\,,\\
	\bm{\sigma}_{p_-p_-}&=\frac{m\omega}{2}\bigl(2\bar{n}+1\bigr)\bigl(\cosh2\eta-\cos\phi\,\sinh2\eta\bigr)\,,\\
	\bm{\sigma}_{\chi_+p_+}&=-\frac{1}{2}\,\bigl(2\bar{n}+1\bigr)\sin\phi\,\sinh2\eta\,,\\
	\bm{\sigma}_{\chi_-p_-}&=+\frac{1}{2}\,\bigl(2\bar{n}+1\bigr)\sin\phi\,\sinh2\eta\,,\label{E:ikgbidehs10}
\end{align}
and the others being zero. The modes $\chi_{\pm}$ are completely decoupled, as can be seen from
\begin{align}
	\langle\bigl\{\chi_{+},\,\chi_{-}\bigr\}\rangle&=0\,,&\langle\bigl\{p_{+},\,p_{-}\bigr\}\rangle&=0\,,
\end{align}
and then the covariance matrix has the form
\begin{equation}\label{E:bskads}
	\bm{\sigma}_{\pm}=\begin{pmatrix}\sigma_{\chi_{+}\chi_{+}} &\sigma_{\chi_{+}p_{+}} &0 &0\\[4pt] \sigma_{\chi_{+}p_{+}} &\sigma_{p_{+}p_{+}} &0 &0\\[4pt] 0 &0 &\sigma_{\chi_{-}\chi_{-}} &\sigma_{\chi_{-}p_{-}}\\[4pt] 0 &0 &\sigma_{\chi_{-}p_{-}} &\sigma_{p_{-}p_{-}}	\end{pmatrix}\,.
\end{equation}
This form is of particular use because it has the same structure as the one in \eqref{E:pworihfbd}.

In this special setting, we find the smaller $\lambda^{\textsc{pt}}_{<}$ of the symplectic eigenvalues of the partially transposed covariance matrix $\bm{\sigma}^{\textsc{pt}}$ of the canonical variables $\bm{Z}=(\chi_{1}, p_{1}, \chi_{2}, p_{2})^{T}$ is given by
\begin{equation}\label{E:kgfbdf}
	\lambda^{\textsc{pt}}_{\gtrless}=e^{\pm2\eta}\,\bigl(\bar{n}+\frac{1}{2}\bigr)\,.
\end{equation}
The entanglement occurs when $\lambda^{\textsc{pt}}_{<}<1/2$
\begin{align}
	e^{-2\eta}\,\bigl(\bar{n}+\frac{1}{2}\bigr)&<\frac{1}{2}\,,&&\Rightarrow&\eta&>\frac{1}{2}\,\ln\bigl(2\bar{n}+1\bigr)=\frac{1}{2}\,\ln\coth\frac{\beta\omega}{2}\,.
\end{align}
In comparison, for the normal modes $\bm{Z}_{\pm}=(\chi_{+}, p_{+}, \chi_{-}, p_{-})^{T}$, the symplectic eigenvalues of the partially transposed covariance matrix $\bm{\sigma}_{\pm}^{\textsc{pt}}$ in this basis are fully degenerate
\begin{equation}\label{E:dgkbs2}
	\bigl(\bar{n}_{1}+\frac{1}{2}\bigr)^{\frac{1}{2}}\bigl(\bar{n}_{2}+\frac{1}{2}\bigr)^{\frac{1}{2}}\,,
\end{equation}
and in this basis, the state is always separable.

At first sight, these results may look dubious because we expect that the entanglement should be a symplectic invariant, independent of the coordinate systems. Indeed, the canonical pair $\bm{Z}$ and the normal modes $\bm{Z}_\pm$ are related by a symplectic transformation. Nonetheless, the negativity is defined via partial transposition, which has assumed a chosen partition a priori, and the partial transposition is not symplectic. Thus the partially transposed pair $\bm{\sigma}^{\textsc{pt}}$ and $\bm{\sigma}_{\pm}^{\textsc{pt}}$ are not related by a symplectic transformation, and their symplectic eigenvalues are in general not identical, as explicitly shown in \eqref{E:kgfbdf} and \eqref{E:dgkbs2}.

\section{Caldeira-Leggett limit for unstable dynamics of an open system}\label{S:gbdghf}

In the high temperature regime, {under} the Caldeira-Leggett approximation, the Hadamard function of the private bath of oscillator $i$ is approximately given by
\begin{equation}
	e^{2}G_{H,0}^{(\xi_{i})}(s-s')=\frac{4m\gamma}{\beta_{i}}\,\delta(s-s')+\text{high-frequency mode contribution}\,.
\end{equation}
If we ignore the high-frequency contribution, this approximation leads to
\begin{equation}
	P_{\xi_{i}}(t)=\frac{2\gamma}{\beta_{i}}\,,
\end{equation}
which is the expression used in \eqref{E:kfgjbdfg} when $\gamma_i=\gamma$ for both private baths. That is, the noise power is constant at all times. Here we look more closely at the validity of this approximation.

In principle, the Hadamard function is ill defined without a proper cutoff $\Lambda$ in its integral representation
\begin{align}\label{E:fgjhjshdds}
	G_{H,0}^{(\xi_{i})}(\tau)=\int_{0}^{\Lambda}\!\frac{d\kappa}{2\pi}\;\frac{\kappa}{4\pi}\,\coth\frac{\beta\kappa}{2}\,\bigl(e^{-i\kappa\tau}+e^{+i\kappa\tau}\bigr)
\end{align}
because the integrand grows linearly with $\kappa$ as $\kappa\to\infty$. Hence we need to be more careful when we take the high-temperature limit. In addition, the high-temperature approximation of $\coth\dfrac{\beta\kappa}{2}$ is given by
\begin{align}
	\coth\frac{\beta\kappa}{2}\simeq\frac{2}{\beta\kappa}+\frac{\beta\kappa}{6}+\cdots\label{E:uiweesbf}
\end{align}
when $\beta\kappa\ll1$. The higher-order terms on the righthand side tend to introduce stronger dependence on the cutoff $\Lambda$.

Suppose we choose a $\omega_{c}$ such that $\beta\omega_{c}\ll1$ and divide the frequency spectrum $[0,\Lambda]$ into two intervals, $[0,\omega_{c}]$ and $[\omega_{c},\Lambda]$, and apply different approximation schemes {in each interval}. Thus we can write \eqref{E:fgjhjshdds} into
\begin{align}\label{E:qzjshdf}
	G_{H,0}^{(\xi_{i})}(\tau)&=\int_{0}^{\omega_{c}}\!\frac{d\kappa}{2\pi}\;\frac{\kappa}{4\pi}\,\Bigl[\frac{2}{\beta\kappa}+\frac{\beta\kappa}{6}+\cdots\Bigr]\,\bigl(e^{-i\kappa\tau}+e^{+i\kappa\tau}\bigr)+\int_{\omega_{c}}^{\Lambda}\!\frac{d\kappa}{2\pi}\;\frac{\kappa}{4\pi}\,\bigl(e^{-i\kappa\tau}+e^{+i\kappa\tau}\bigr)\notag\\
	&=\frac{\sin\omega_{c}\tau}{2\pi^{2}\beta\tau}+\frac{2\beta\omega_{c}\tau\,\cos\omega_{c}\tau+\beta(\omega_{c}^{2}\tau^{2})\sin\omega_{c}\tau}{24\pi^{2}\tau^{3}}\notag\\
 &\qquad\qquad\qquad\qquad+\frac{\cos\Lambda\tau-\cos\omega_{c}\tau+\Lambda\tau\,\sin\Lambda\tau-\omega_{c}\tau\,\sin\omega_{c}\tau}{4\pi^{2}\tau^{2}}+\cdots
\end{align}
The first two terms are the high-temperature contributions, while the third term results from the zero-point fluctuations, which is important in the high-frequency modes.

In the first term, for sufficiently small $\beta$, the parameter $\omega_{c}$ can be allowed to be very large such that the first term approximately gives
\begin{equation}\label{E:fkbgdbvds}
	\frac{\sin\omega_{c}\tau}{2\pi^{2}\beta\tau}\simeq\frac{1}{2\pi\beta}\,\delta(\tau)\,,
\end{equation}
due to the formula
\begin{equation}
	\lim_{\epsilon\to0}\frac{\sin(x/\epsilon)}{\pi x}=\delta(x)\,.
\end{equation}
This is the result of the Caldeira-Leggett limit. In deriving \eqref{E:fkbgdbvds}, we do not put any restriction on $\tau$ as long as $\omega_c\tau\gg1$. 

The second term in \eqref{E:qzjshdf} can be broken down to
\begin{align}\label{E:fkuiwsdfd}
	\frac{\beta}{\tau}\times\frac{\beta\omega_{c}}{12\pi^{2}\beta\tau}\,\cos\omega_{c}\tau+(\beta\omega_{c})^{2}\,\frac{1}{24\pi^{2}\beta\tau}\,\sin\omega_{c}\tau\,,
\end{align}
the second of which is apparently much smaller than \eqref{E:fkbgdbvds} in the high-temperature regime, while the first of which becomes subleading, relative to the second contribution in \eqref{E:fkuiwsdfd}, only when $\omega_{c}\tau>1$. We might be led to argue that up to the order in the series expansion \eqref{E:uiweesbf}, this term may have a non-negligible contribution at early time $\tau\ll\beta$, depending on the factor
\begin{equation}
	\frac{\beta}{\tau}\times(\beta\omega_{c})\gtrless1\,.
\end{equation}
It turns out that this argument is incorrect because the second contribution in \eqref{E:fkuiwsdfd} is not taken into count yet. Once we put them together, we find that at early times \eqref{E:fkuiwsdfd} gives
\begin{equation}
	\frac{\beta\omega_{c}^{3}}{72\pi^{2}}\,,
\end{equation}
the order $(\beta\omega_{c})^{2}$ smaller than the corresponding contribution from \eqref{E:fkbgdbvds} at early times. Thus the second term in \eqref{E:qzjshdf} can be safely ignored in the high-temperature limit.

The third term in \eqref{E:qzjshdf} originates from the vacuum fluctuations of the high-frequency modes, and the cutoff typically marks the highest possible frequency of the modes compatible with the model or the underlying theory. Thus, we actually have implicitly required
\begin{equation}
	\beta\Lambda\gg1\,,
\end{equation}
and $\Lambda\gg\omega_{c}$ for the consistency's sake. The important contribution comes from
\begin{equation}
	\frac{\Lambda\tau\,\sin\Lambda\tau}{4\pi^{2}\tau^{2}}=(\beta\Lambda)\times\frac{\sin\Lambda\tau}{4\pi^{2}\beta\tau^{2}}\,.
\end{equation}
It is typically much larger than \eqref{E:fkbgdbvds}. In addition, at early time this contribution behaves like
\begin{equation}
	\frac{\Lambda^{2}}{8\pi^{2}}\,,
\end{equation}
a potentially vary large level, compared to
\begin{equation}
	\frac{\omega_{c}}{2\pi^{2}\beta}
\end{equation}
from \eqref{E:fkbgdbvds} with a ratio
\begin{equation}
	(\beta\Lambda)\times\frac{\Lambda}{\omega_{c}}\gg\gg1\,.
\end{equation}
Thus from the analysis of the Hadamard function \eqref{E:fkuiwsdfd}, we find the inconsistency of the result from the Caldeira-Leggett limit at high temperature.

\begin{figure}
\centering
    \scalebox{0.45}{\includegraphics{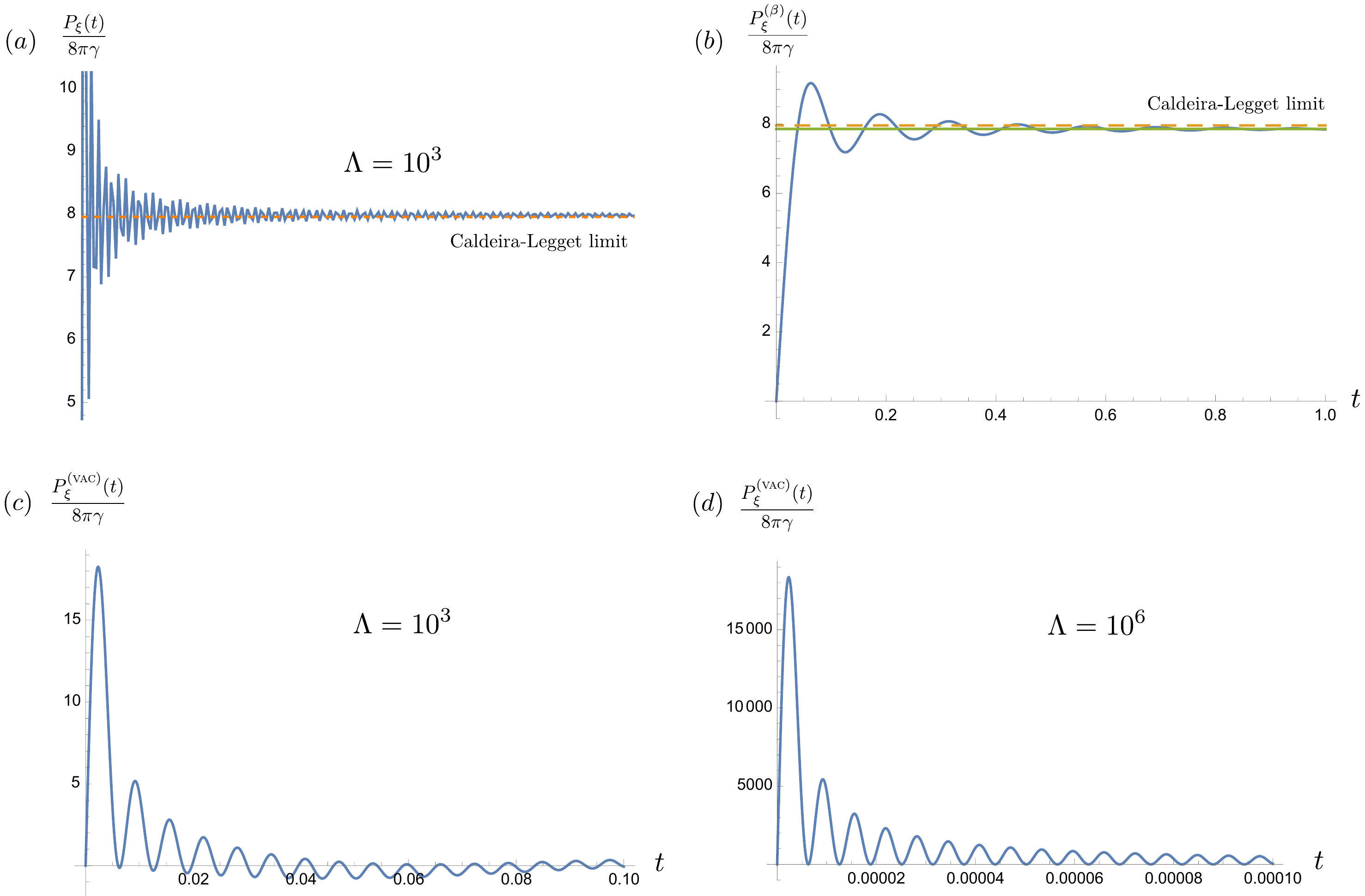}}
    \caption{The early time behavior of (a) $P_{\xi}(t)$, (b) the finite temperature contribution $P_{\xi}^{(\beta)}(t)$ in \eqref{E:csdersjd} and (c)\&(d) the vacuum contribution $P_{\xi}^{(\textsc{vac})}(t)$ in \eqref{E:csdersjd}. Plot (a) is generated without applying the high temperature approximation. It looks quite noisy due to the ripples caused by the cutoff. The curve oscillates at frequency $\Lambda$ but with an exponentially decreasing amplitude. The Caldeira-Leggett limit is denoted by the orange dashed line in (a) and (b). The jolt in $P_{\xi}^{(\textsc{vac})}(t)$ has a magnitude proportional to $\Lambda$ within the duration proportional to $\Lambda^{-1}$. Here we choose $\gamma=0.5$, $\omega_{\textsc{p}}=1$, $\omega_{c}=50$, and $\beta=0.01$.}\label{Fi:CL}
\end{figure}

However, when the Hadamard function is part of the integrand of a time integral, will the vacuum contribution still dominate over the the Caldeira-Leggett term in the high temperature limit? Let us consider the simplest setting, a {single} Brownian oscillator {coupled to the thermal bath}~\cite{HH15AOP,HC18}, and examine the corresponding noise power $P_{\xi}$,
\begin{align}\label{E:eiufgsjdw}
	P_{\xi}(t)&=\frac{e^{2}}{m}\int_{0}^{t}\!ds\;\dot{d}_{2}(t-s)\,G_{H,0}^{(\xi)}(t-s)\notag\\
				&=\frac{e^{2}}{m}\biggl\{\int_{0}^{\omega_{c}}\!\frac{d\kappa}{2\pi}\;\frac{\kappa}{4\pi}\frac{2}{\beta\kappa}+\int_{\omega_{c}}^{\Lambda}\!\frac{d\kappa}{2\pi}\;\frac{\kappa}{4\pi}\biggr\}\int_{0}^{t}\!ds\;\dot{d}_{2}(t-s)\,\bigl[e^{-i\kappa(t-s)}+e^{+i\kappa(t-s)}\bigr]\,,
\end{align}
with $d_{2}(t)$ of the damped harmonic oscillator given by
\begin{equation}
	d_{2}(t)=\frac{1}{\Omega}\,e^{-\gamma t}\,\sin\Omega t\,,
\end{equation}
where $\Omega^{2}=\omega_{\textsc{p}}^{2}-\gamma^{2}$. At late times, in the high temperature limit, we find
\begin{align}\label{E:csdersjd}
	P_{\xi}(\infty)&=\biggl[\frac{2\pi}{\pi\beta}\Bigl(\cot^{-1}\frac{\gamma}{\omega_{c}-\Omega}+\cot^{-1}\frac{\gamma}{\omega_{c}+\Omega}\Bigr)-\frac{2\gamma^{2}}{\pi\beta\Omega}\tanh^{-1}\frac{2\Omega\omega_{c}}{\Omega^{2}+\gamma^{2}+\omega_{c}^{2}}\biggr]\notag\\
	&\qquad\qquad+\frac{\gamma^{2}}{\pi}\,\ln\frac{(\Lambda^{2}+\gamma^{2})^{2}-2(\Lambda^{2}-\gamma^{2})\Omega^{2}+\Omega^{4}}{(\omega_{c}^{2}+\gamma^{2})^{2}-2(\omega_{c}^{2}-\gamma^{2})\Omega^{2}+\Omega^{4}}\,.
\end{align}
The first term in \eqref{E:csdersjd} gives the Caldeira-Leggett limit $\frac{2\gamma}{\beta}$ to very high accuracy, and the second term is the vacuum, cutoff-dependent contribution. Two terms can be comparable when $\Lambda\sim\omega_{c}\exp(\frac{\pi^{2}}{\beta\gamma})$, which practically is a tremendously high frequency where {the vacuum contribution is ignorable}.  For the stable motion consider here, at high temperature, the Caldeira-Leggett limit gives a sufficiently reliable result at late times.

On the other hand, at early times, deviation from the Caldeira-Leggett limit shows up. In Fig.~\ref{Fi:CL}, we see the finite-temperature contribution $P_{\xi}^{(\beta)}$ falls to zero as $t\to 0$ and the vacuum contribution $P_{\xi}^{(\textsc{vac})}$ has a huge jolt with the magnitude proportional to $\Lambda$ for a duration proportional to $\Lambda^{-1}$. These imply that the Caldeira-Leggett approximation is best suited for high-temperatures in an equilibrium setting, not for seeking early time behavior in a nonequilibrium setting.

It is interesting to compare with the case of the damped inverted oscillator. The corresponding $d_{2}(t)$ has the form
\begin{equation}
	d_{2}(t)=\frac{1}{\Omega}\,e^{-\gamma t}\,\sinh\Omega t\,.
\end{equation}
Here $\Omega^{2}=\omega_{\textsc{p}}^{2}+\gamma^{2}$. Apparently it grows exponentially fast. We find
\begin{align}
	&\quad\int_{0}^{t}\!ds\;d_{2}(t-s)\Bigl[e^{-i\kappa(t-s)}+e^{+i\kappa(t-s)}\Bigr]\notag\\
	&=\frac{1}{\Omega[\kappa^{2}+(\Omega-\gamma)^{2}][\kappa^{2}+(\Omega+\gamma)^{2}]}\\
	&\qquad\qquad\times\biggl\{4\gamma\Omega\,\kappa^{2}+2\Omega\kappa\,e^{-\gamma t}\bigl[-2\gamma\kappa\,\cos\kappa t+\bigl(\kappa^{2}+\omega_{\textsc{p}}^{2}\bigr)\,\sin\kappa t\bigr]\cosh\Omega t\biggr.\notag\\
	&\qquad\qquad\qquad\qquad+\biggl.e^{-\gamma t}\bigl[2\bigl(\omega_{\textsc{p}}^{4}+\bigl(\Omega^{2}+\gamma^{2}\bigr)\kappa^{2}\bigr)\cos\kappa t-2\gamma\kappa\,\bigl(\kappa^{2}-\omega_{\textsc{p}}^{2}\bigr)\sin\kappa t\bigr]\sinh\Omega t\biggr\}\,,\notag
\end{align}
This is still exponentially increasing with $t$ because $\Omega>\gamma$, but oscillates with the frequency $\kappa$. We follow the same strategy and divide the frequency band $\kappa\in[0,\Lambda]$ into two intervals, such that we write $P_{\xi}(t)$ as
\begin{align}\label{E:dkdbcxvbse}
	P_{\xi}(t)&=\frac{e^{2}}{m}\biggl\{\int_{0}^{\omega_{c}}\!\frac{d\kappa}{2\pi}\;\frac{\kappa}{4\pi}\frac{2}{\beta\kappa}+\int_{\omega_{c}}^{\Lambda}\!\frac{d\kappa}{2\pi}\;\frac{\kappa}{4\pi}\biggr\}\int_{0}^{t}\!ds\;\dot{d}_{2}(t-s)\,\bigl[e^{-i\kappa(t-s)}+e^{+i\kappa(t-s)}\bigr]\,.
\end{align}
The first integral in \eqref{E:dkdbcxvbse}, corresponding to the finite-temperature contribution, gives
\begin{align}
	&\frac{2\gamma}{\pi\beta\Omega}\biggl[\bigl(\Omega+\gamma\bigr)\cot^{-1}\frac{\Omega+\gamma}{\omega_{c}}-\bigl(\Omega-\gamma\bigr)\cot^{-1}\frac{\Omega-\gamma}{\omega_{c}}\biggr]\\
	&\qquad\qquad\qquad\qquad-i\,\frac{\gamma}{\pi\beta\Omega}\biggl[\frac{\Omega-\gamma}{\Omega-\gamma+i\omega_{c}}\frac{e^{(\Omega-\gamma)t+i\omega_{c}t}}{t}-\frac{\Omega-\gamma}{\Omega-\gamma-i\omega_{c}}\frac{e^{(\Omega-\gamma)t-i\omega_{c}t}}{t}+\cdots\biggr]\,,\notag
\end{align}
at late times. The first term will give the Caldeira-Leggett limit, but the second term oscillates with an exponentially increasing amplitude. Thus its contribution cannot be simply ignored. If we can make sense out of the averaging of these oscillations, the Caldeira-Leggett limit emerges. The second integral in \eqref{E:dkdbcxvbse} gives the vacuum contributions, which introduce the high-frequency ripples into $P_{\xi}(t)$, 
\begin{align}
	&\frac{\gamma^{2}}{\pi}\ln\frac{(\Lambda^{2}+\Omega^{2})^{2}+2\gamma^{2}(\Lambda^{2}-\Omega^{2})+\gamma^{4}}{(\omega_{c}^{2}+\Omega^{2})^{2}+2\gamma^{2}(\omega_{c}^{2}-\Omega^{2})+\gamma^{4}}+\frac{2\gamma}{\pi\Omega}\,e^{-\gamma t}\,\frac{-\Omega\,\cosh\Omega t+\gamma\,\sinh\Omega t}{t}\,\cos\Lambda t\notag\\
	&\qquad\qquad\qquad\qquad+\frac{\gamma(\Omega-\gamma)^{2}}{\pi\Omega[\Lambda^{2}+(\Omega-\gamma)^{2}]}\frac{(\Omega-\gamma)\cos\Lambda t+\Lambda\,\sin\Lambda t}{t}\notag\\
	&\qquad\qquad\qquad\qquad+\frac{\gamma(\Omega-\gamma)\omega_{c}}{\pi\Omega[\omega_{c}^{2}+(\Omega-\gamma)^{2}]}\frac{\omega_{c}\cos\omega_{c} t-(\Omega-\gamma)\sin\omega_{c} t}{t}+\cdots\,.
\end{align}
This gives a very low noise base at late times due to small $\gamma$ and the appearance of $\Lambda$ in the logarithm. However, it still has an oscillatory contribution with the same exponentially increasing amplitude. Note that our analysis, though providing a quick glance into the contribution of the Caldeira-Leggett term in $P_{\xi}(t)$, inadvertently introduces an artifact because we divide the frequency band into two intervals, in which two distinct approximations are used. The result has a component oscillating with the frequency $\omega_{c}$. This {is} not present in the numerical evaluation of $P_{\xi}$. Actually, $P_{\xi}(t)$ will oscillates roughly about the Caldeira-Leggett limit, at a frequency determined by the cutoff $\Lambda$ due to vacuum fluctuations, but with an exponentially increasing amplitude, a consequence of  the inverted potential. Thus we see for the case of inverted oscillator, the subtlety in applying the Caldeira-Leggett limit in $P_{\xi}(t)$ mainly comes from oscillations with ever increasing amplitude, instead of the contributions from vacuum fluctuations.

\begin{figure}
\centering
    \scalebox{0.6}{\includegraphics{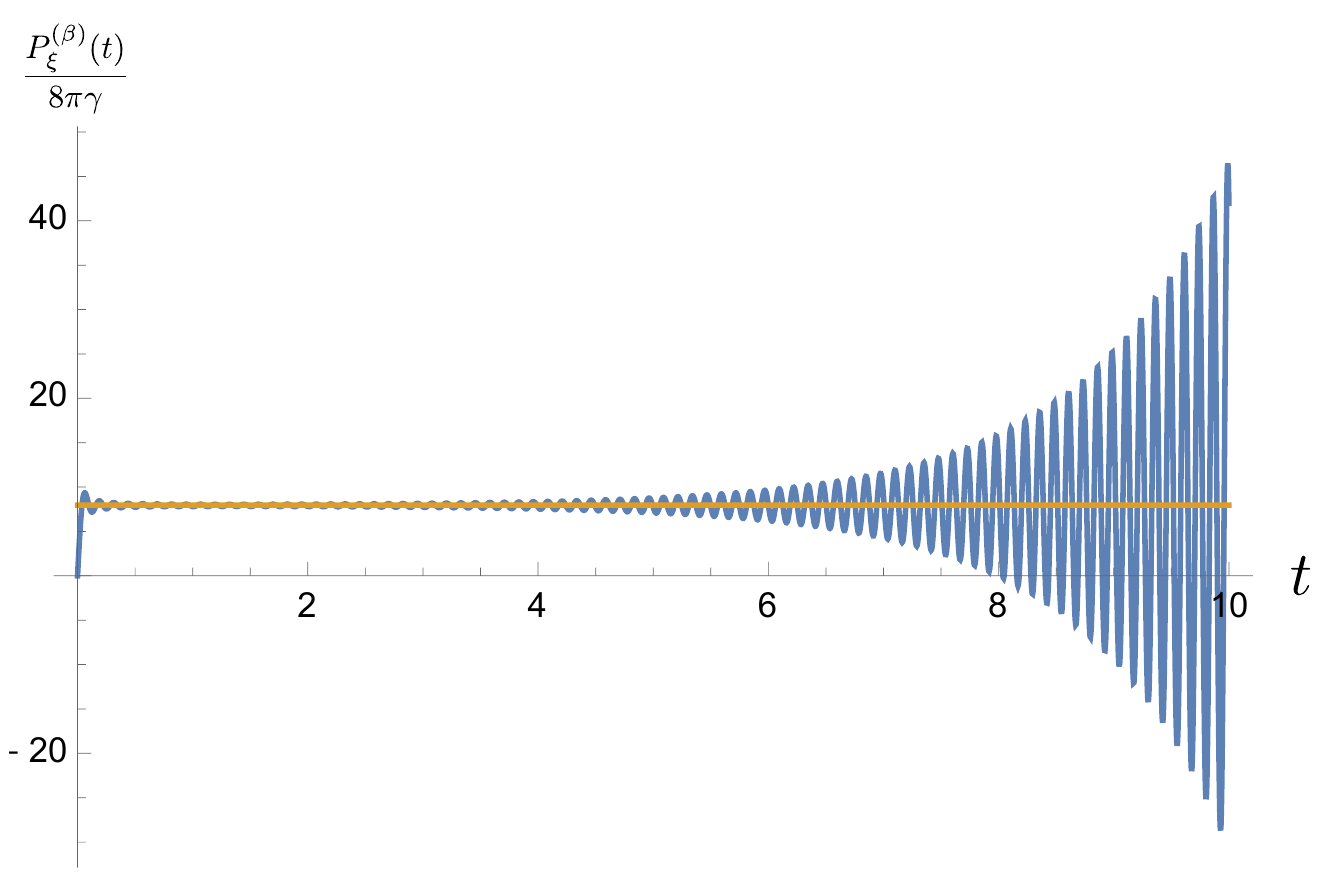}}
    \caption{temporal behavior of the finite temperature contribution $P_{\xi}^{(\beta)}(t)$ of the noise power. The orange dashed line gives the Caldeira-Leggett term. We see eventually $P_{\xi}^{(\beta)}(t)$ oscillates about the Caldeira-Leggett term with an exponentially large amplitude, in contrast to the Brownian case. We choose $\gamma=0.02$, $\omega_{\textsc{p}}=1$, $\omega_{c}=50$, and $\beta=0.01$.}\label{Fi:CL_amp}
\end{figure}

The reason that the Caldeira-Leggett limit gives a seemingly plausible result for $P_{\xi}$ lies in a few features $P_{\xi}$ has. The first is the functional form of $P_{\xi}(t)$, which is solely expressed as an integral of the oscillatory integrand, as seen in \eqref{E:eiufgsjdw}. This implies that $P_{\xi}(t)$ in general is oscillatory and thus is not sign definite. Moreover, to make the integral well defined, a cutoff is introduced to the integration limit. The consequences are that 1) the results tend to oscillate with very high frequency (in particular in the inverted oscillator case, this is the only scale that is related to oscillatory behavior), and 2) there exists a cutoff-dependent contribution. However, the latter in general has the form $\gamma^{2}\ln\Lambda$, so it is much smaller than the finite temperature contribution in the high-temperature regime. On the other hand, the Caldeira-Leggett term gives the non-oscillating component in $P_{\xi}(t)$ at high temperature, so if we could define an {average/smearing} for the amplifying oscillations, then the $P_{\xi}(t)$ would yield the Caldeira-Leggett term pretty accurately.

In comparison, $P_{\gamma}(t)$ by construction is always sign definite, so it always dissipates the energy of the reduced system. Thus, following our earlier discussions, on the average sense, for the inverted oscillator case, the energy gained from falling down the potential per unit time is mostly distributed among the kinetic energy of the reduced system and its frictional loss. The energy exchange via the noise channel barely plays any role. This is the reason why a fluctuation-dissipation relation does not exist,  because $P_{\gamma}$ grows way out of proportion.

\end{document}